%% file: bare_jrnl_new_sample4.tex
\documentclass[lettersize,journal]{IEEEtran}
\usepackage{amsmath,amsfonts}
\usepackage{algorithmic}
\usepackage{algorithm}
\usepackage{array}
\usepackage[caption=false,font=normalsize,labelfont=sf,textfont=sf]{subfig}
\usepackage{textcomp}
\usepackage{stfloats}
\usepackage{url}
\usepackage{verbatim}
\usepackage{graphicx}
\usepackage{cite}
\usepackage{algorithm}
\usepackage{algorithmic}
\usepackage{amssymb}
\usepackage{amsmath}
\usepackage[colorlinks,linkcolor=blue]{hyperref}

\usepackage{booktabs}
\usepackage{tabularx}
\usepackage{colortbl}  
\usepackage{xcolor}
\usepackage{array}   
\definecolor{lightgray}{gray}{0.95}
\definecolor{color3}{gray}{0.95}
\definecolor{rouse}{rgb}{0.981,0.961,0.941}

\begin{document}

\title{Degradation Estimation Recurrent Neural Network with Local and Non-Local Priors for Compressive Spectral Imaging}

\author{Yubo Dong, \textit{Member, IEEE}, Dahua Gao *, \textit{Member, IEEE}, Yuyan Li, \textit{Member, IEEE}, \\ Guangming Shi, \textit{Fellow, IEEE}, and Danhua Liu, \textit{Member, IEEE}  
}

\markboth{Journal of \LaTeX\ Class Files,~Vol.~14, No.~8, August~2021}%
{Shell \MakeLowercase{\textit{et al.}}: A Sample Article Using IEEEtran.cls for IEEE Journals}


\maketitle

\begin{abstract}
In the Coded Aperture Snapshot Spectral Imaging (CASSI) system, deep unfolding networks (DUNs) have demonstrated excellent performance in recovering 3D hyperspectral images (HSIs) from 2D measurements. However,  some noticeable gaps exist between the imaging model used in DUNs and the real CASSI imaging process,  such as the sensing error as well as photon and dark current noise, compromising the accuracy of solving the data subproblem and the prior subproblem in DUNs. To address this issue, we propose a Degradation Estimation Network (DEN) to correct the imaging model used in DUNs by simultaneously estimating the sensing error and the noise level, thereby improving the performance of DUNs. Additionally, we propose an efficient Local and Non-local Transformer (LNLT) to solve the prior subproblem, which not only effectively models local and non-local similarities but also reduces the computational cost of the window-based global Multi-head Self-attention (MSA). Furthermore, we transform the DUN into a Recurrent Neural Network (RNN) by sharing parameters of DNNs across stages, which not only allows DNN to be trained more adequately but also significantly reduces the number of parameters. The proposed DERNN-LNLT achieves state-of-the-art (SOTA) performance with fewer parameters on both simulation and real datasets. Code: \href{https://github.com/ShawnDong98/DERNN-LNLT}{https://github.com/ShawnDong98/DERNN-LNLT}
\end{abstract}


\section{Introduction}

In various fields including remote sensing \cite{rs1, rs2, rs3}, biomedical research \cite{med1, med2}, and industrial inspection \cite{industrial}, hyperspectral imaging technology, known for its ability to simultaneously capture spectral and spatial information, has been widely used.  To acquire hyperspectral images (HSIs), numerous hyperspectral imaging techniques have been developed. Conventional methods include whiskbroom scanning and pushbroom scanning. However, these techniques typically require multiple exposures and complex 2D or 1D scanning mechanisms, leading to slow imaging speed, making them unsuitable for dynamic scenes.

The coded aperture snapshot spectral imaging (CASSI) technique is proposed for capturing snapshot spectral images \cite{arce2013compressive, wagadarikar2008single, wang2015high}. It employs a coded aperture in conjunction with a dispersive prism to modulate the spectral information of a scene, enabling the CASSI to acquire a multiplexed 2D projection of the 3D HSI for real-time spectral data acquisition. However, a fundamental challenge is posed in reconstructing the high-fidelity 3D HSI from the 2D measurement.

DUNs \cite{gap-net, admm-net, dgsmp, DPIR, herosnet, dauhst, rdluf_mixs2} have shown superior performance in HSI reconstruction by alternately solving a data subproblem and a prior subproblem. However, there are some noticeable gaps between the imaging model used in DUNs and the real CASSI imaging process, such as the sensing error between the sensing matrix and the real degradation process, as well as photon and dark current noise. These gaps compromise the reconstruction performance of DUNs. The degradation matrix plays a pivotal role in the data subproblem. Recent works have started to deal with the sensing error. For instance, \cite{dgsmp} uses CNNs to estimate the degradation matrix, while \cite{rdluf_mixs2} obtains the degradation matrix through residual learning with reference to the sensing matrix. Furthermore, some works have begun considering the role of noise level. \cite{DPIR} suggests that the prior subproblem can be viewed as a denoising problem at a specific noise level, and it uses a predefined noise level for different stages. Additionally, \cite{dauhst} estimates the noise level from the CASSI system. However, considering either the sensing error or the noise level leads to an inaccurate imaging model, compromising the accuracy of solving either the data subproblem or the prior subproblem in DUNs.

To address these issues, we propose a Degradation Estimation Network (DEN) for DUNs, which simultaneously estimates the degradation matrix and the noise level by residual learning with reference to the sensing matrix. The estimated parameters of the DEN can significantly enhance the accuracy of solving both the data subproblem and the prior subproblem. Additionally, we use the residual between the sensing matrix and the degradation matrix as an intermediate feature map to predict the noise level. This approach allows for the sharing of convolutional neural networks (CNNs) within the DEN, thereby reducing the number of required parameters.

In the prior subproblem, a denoiser is trained to implicitly represent the prior term as a denoising problem. It has been demonstrated that both local and non-local priors exist in HSIs \cite{wang2016adaptive, zhang2019computational}.  CNNs \cite{l-net, tsa-net, dgsmp} and window-based local Transformers \cite{swin} have achieved success in modeling local similarities. However, their limited receptive field poses a challenge in identifying non-local similarities. On the other hand, window-based global Transformers \cite{vit} can capture non-local similarities across windows, but their implicit modeling of spatial features within a window often leads to the loss of important local textures and details. In addition, the computational cost of the window-based global MSA is unaffordable in HSIs. It increases linearly with the number of spectral bands for projection and quadratically with spatial size for self-attention. Recent works \cite{maxim, dauhst} model non-local similarity by shuffling/permuting pixels, allowing long-range pixels to be positioned within a local window.  However, these methods, focusing only on position-specific pixel-level non-local similarities, often neglect key relevant non-local pixels and long-range patch-level similarities \cite{nlm}, leading to inadequate reconstructions and missing object-level relationships.

\begin{figure}[t]
    \begin{center}
        \begin{tabular}[t]{c} \hspace{-3.8mm} 
            \includegraphics[height=0.375\textwidth, width=0.5\textwidth]{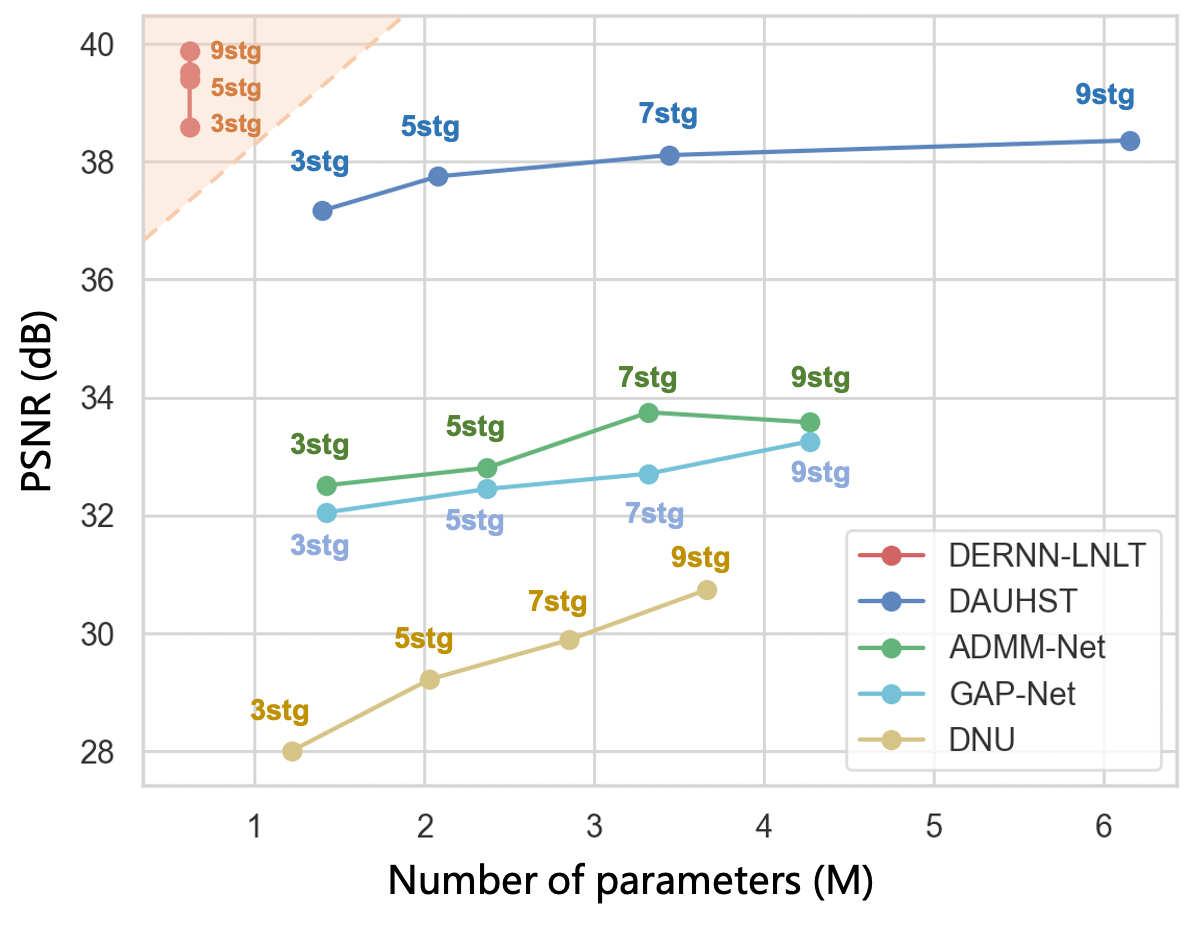}
        \end{tabular}
    \end{center}
    \caption{\small Comparison of PSNR-Params with previous HSI DUNs. The PSNR (in dB) is plotted on the vertical axis, while the number of parameters is represented on the horizontal axis. The proposed DERNN-LNLT outperforms the previous DUNs while requiring much fewer parameters.}
    \label{fig:teaser}
\end{figure}

To address these issues, we propose the Local and Non-Local Transformer (LNLT), which leverages the advantages of both local and global Transformer.  In contrast to window-based global MSA that uses a fixed window size, we fix the number of windows. This not only enables the computational cost of the self-attention to increase linearly with the spatial size but also allows the window of the global MSA to cover multiple windows of the local MSA, enabling better modeling of non-local similarities across windows. Furthermore, we alter the direction of projection in the global MSA to the channel dimension to reduce the computational cost caused by the large number of spectral bands in HSIs. This adjustment ensures that the computational cost for projection remains the same for any window size. By leveraging the advantages of both local and global MSA, the LNLT ensures that every pixel contributes to the receptive fields to model non-local similarities and enables the modeling of object-level non-local relationships.

DUNs achieve superior performance by unfolding model-based methods into a cascade of deep neural networks (DNNs). However, the cascading of multi-stage DNNs brings three primary challenges. Firstly, the isolation of DNNs across different stages results in the DNN at each stage struggling to receive adequate training, resulting in a significant number of neurons remaining non-activated, thereby leading to suboptimal reconstruction. Secondly, the multi-stage cascading architecture leads to unstable training and optimization \cite{mprnet}. Lastly, stacking multiple DNNs results in a substantial increase in parameters. Previous methods \cite{mprnet, dgunet, dgsmp, rdluf_mixs2} try to address the first two challenges by introducing intricate connections or interactions across stages. However, these methods not only fail to address the issue of excessive parameters but also compound the problem by adding extra modules for processing and fusing features, consequently increasing both the number of parameters and memory usage.

We address these three issues simultaneously in a simple way, without relying on intricate connections or interactions across stages. By sharing parameters across different stages of DNNs, we transform the DNN into an RNN. This approach enables the DNN to learn denoising images with various noise levels, which can be regarded as multi-task learning. The multi-task learning forces the DNN to possess denoising capabilities for different noise levels, thereby necessitating the activation of more neurons and enhancing the generalization of the DNN. The RNN inherently allows for the flow of information across different stages, enabling stable training without intricate connections and interactions. Lastly, parameter sharing in RNN dramatically reduces the required parameters and memory consumption compared to DNNs.

Our contributions can be summarized as follows:
\begin{itemize}
    \item We propose a Degradation Estimation Network (DEN) to correct the gaps between the imaging model used in DUNs and the real CASSI imaging process,  which improves the accuracy of solving the data subproblem and the prior subproblem of DUNs.
    \item The efficient Local and Non-Local Transformer (LNLT) is proposed to effectively exploit both local and non-local priors. The crafted design of LNLT not only significantly reduces the computational cost of window-based global MSA, but also ensures that every pixel contributes to the receptive fields to model non-local similarities and enables the modeling of object-level non-local relationships.
    \item We transform the DUN into an RNN by sharing parameters of DNNs across stages. The shared DNN activates more neurons, thereby enhancing its generalization and robustness. At the same time, the design of RNN stabilizes the training process and dramatically reduces parameters.
    \item The proposed method, named DERNN-LNLT,  achieves SOTA performance with fewer parameters as shown in Fig. \ref{fig:teaser}.
\end{itemize}

\section{Related Works}

\subsection{Deep Unfolding Networks}

Model-based methods often adopt a Bayesian perspective, framing the HSI reconstruction as a MAP optimization. Commonly used algorithms include PGD \cite{pgd}, ADMM \cite{admm}, and HQS \cite{hqs}. These methods typically separate data fidelity and prior terms, leading to an iterative process alternating between solving a data subproblem and a prior subproblem. However, model-based methods suffer from the poor representativeness of hand-crafted priors, resulting in limited reconstruction quality. To leverage the representation capabilities of DNNs, deep unfolding networks (DUNs) unfold the model-based methods as a cascade of DNNs \cite{meinhardt2017learning, ryu2019plug, yuan2020plug, zhang2017learning, zhang2019deep}. DUNs have achieved success in HSI reconstruction. Nevertheless, the independent DNNs in DUNs not only result in numerous parameters and significant memory costs but also lead to suboptimal reconstruction performance due to inadequate training of the DNN at each stage.

\subsection{Degradation-related Parameters in DUNs}

 Some noticeable gaps exist between the imaging model used in DUNs and the real CASSI imaging process,   compromising the accuracy of solving the data subproblem and the prior subproblem in DUNs, such as the sensing error between the sensing matrix and the real degradation process, as well as photon and dark current noise.

In the data subproblem, the degradation matrix is crucial for solving it. \cite{dauhst, DPIR, dgsmp-pami} employ the sensing matrix as the degradation matrix. \cite{dgsmp, herosnet} employ a DNN to directly estimate the degradation matrix. However, for the former, the captured sensing matrix during calibration may not accurately reflect the degradation process when capturing real scenes, as it can be affected by variations in exposure time, light intensity, and device errors. For the latter, it is challenging to directly model the degradation process. \cite{rdluf_mixs2} proposes estimating the degradation matrix through residual learning with reference to the sensing matrix to address these issues, integrating the strengths of these two methods.

In the prior subproblem, it can be treated as a denoising problem on an image with a specific noise level \cite{DPIR, dauhst}. Thus, the noise level is critical for solving the denoising problem. However, there is no pre-given noise level in the reconstruction task. \cite{dauhst} estimates the degradation pattern and noise level from the CASSI system. 

Previous works consider either the sensing error or the noise level, without addressing these degradation-related parameters within a unified framework. Consequently, either the accuracy of solving the data subproblem or the performance of solving the prior subproblem is compromised.

\subsection{Methods for Exploiting HSI Priors}

In DUNs, the prior subproblem is implicitly expressed as a denoising problem by employing a trainable denoiser. Therefore, it is significant to design a suitable denoiser to exploit priors in HSIs. Previous works \cite{wang2016adaptive, zhang2019computational} have demonstrated the existence of both local and non-local priors in HSIs. CNN-based methods \cite{l-net, tsa-net} showcase robust capabilities in modeling local similarities. However, while they excel in certain tasks, CNN-based techniques face limitations in identifying non-local similarities due to their inductive biases. To address these issues, Transformer-based approaches \cite{mst, cst} have emerged, leveraging multi-head self-attention (MSA) to capture non-local dependencies in HSIs. Nonetheless, these methods might lead to a loss of critical local features, such as textures and structures, which play a vital role in generating high-quality HSI images. \cite{dauhst} proposes a Half-Shuffle Multi-head self-attention (HS-MSA) that splits attention heads into a local branch and a non-local branch. In this method, modeling non-local similarity is achieved by shuffling pixels, allowing long-range pixels to be positioned within a local window. However, this approach can only model non-local similarities of position-specific pixels, potentially ignoring highly correlated non-local pixels.  Additionally, modeling non-local similarities at the pixel-level may neglect some object-level non-local similarities. Consequently, designing a network that effectively exploits both local and patch-level non-local priors in HSIs holds significant importance.

\begin{figure}[]
    \centering
    \includegraphics[width=0.8\linewidth]{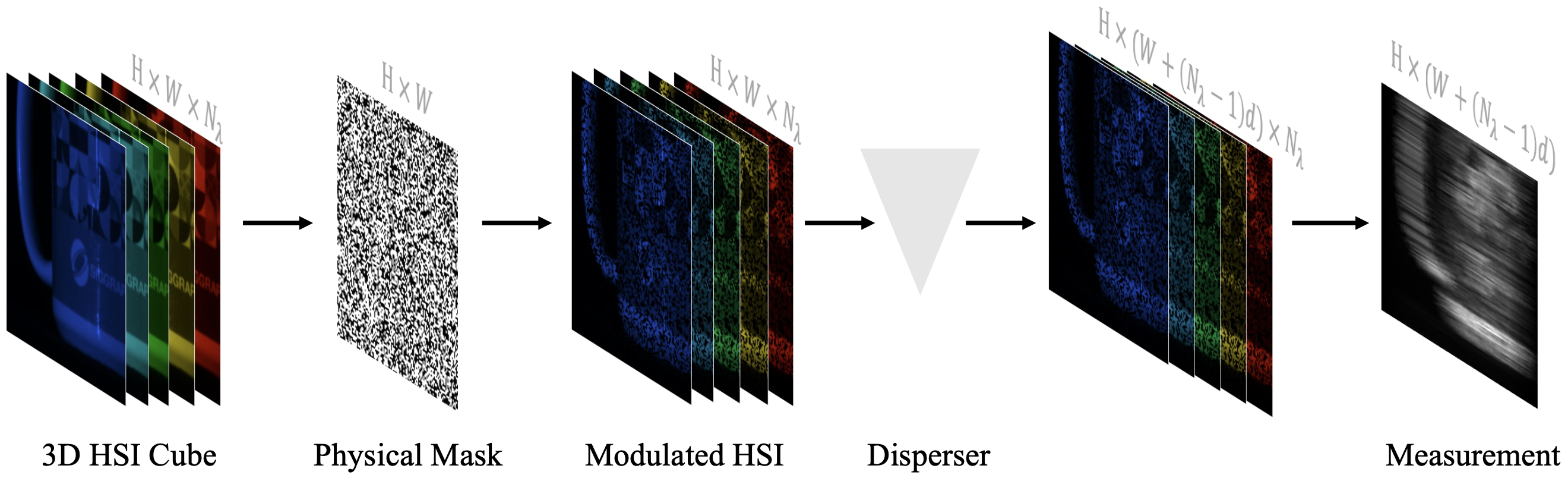}
    \caption{\small Schema of the CASSI system. The 3D HSI cube is modulated by the physical mask and the disperser and ultimately compressed into a 2D measurement. } 
    \label{fig:cassi}
\end{figure}

\section{Reconstruction Problem in CASSI System}

In CASSI \cite{arce2013compressive}, a 3D HSI cube is initially modulated by a physical mask. Subsequently, different wavelengths of the HSI undergo dispersion in the width dimension through a disperser. Finally, the dispersed wavelengths are captured by a 2D imaging sensor and compressed into a single 2D measurement. Fig. \ref{fig:cassi} illustrates the imaging process of the CASSI system.

Consider an HSI signal tensor $X \in \mathbb{R}^{H \times W \times N_\lambda}$ and a physical mask tensor $M \in \mathbb{R}^{H \times W}$. Here, $N_\lambda$ denotes the number of wavelengths. The modulated image at the $n_\lambda$-th wavelength can be represented as follows:
\begin{equation}
    X_{n_\lambda}' = M \odot X_{n_\lambda},
    \label{eq: mask}
\end{equation}
where $\odot$ denotes the element-wise product. Subsequently, the spatially modulated HSI $X'$ is spectrally dispersed by the disperser, which can be expressed as: 
\begin{equation}
    X''(h, w + d_{n_\lambda}, n_\lambda) = X'(h, w, n_\lambda),
    \label{eq: disperse}
\end{equation}
where $X'' \in \mathbb{R}^{H \times (W + d_{N_\lambda}) \times N_\lambda}$, $d_{n_\lambda}$ denotes the dispersed distance of the $n_\lambda$-th wavelength. At last, the imaging sensor captures the dispersed HSI into a 2D measurement, which can be formulated as follows:
\begin{equation}
    Y = \sum_{n_\lambda = 1}^{N_\lambda} X_{n_\lambda}'' + G,
    \label{eq: ill-posed inverse}
\end{equation}
where $Y, G \in \mathbb{R}^{H \times (W + d_{N_\lambda})}$, and $G$ represents the additive noise. The spatial dimensions of the signal are increased due to the dispersion process, while the spectral dimension is compressed by the 2D imaging process.

Mathematically, Eq. (\ref{eq: ill-posed inverse}) is equivalent to the following matrix-vector form:
\begin{equation}
    y = \Phi x + n,
    \label{eq: matrix-vector}
\end{equation}
where $x \in \mathbb{R}^{HW'N_{\lambda}}, W' = W + d_{N_\lambda}$ is the shifted original HSI, $y \in \mathbb{R}^{HW'}$ is the measurement, $\Phi \in \mathbb{R}^{HW' \times HW'N_{\lambda}}$ is the sensing matrix, typically treated as the shifted mask. HSI reconstruction aims to restore the high-quality image $x$ from the measurement $y$, which is typically an ill-posed problem.

\section{The Proposed Method}

\begin{figure*}[!htb]
    \centering
    \includegraphics[width=\linewidth, height=0.9\linewidth]{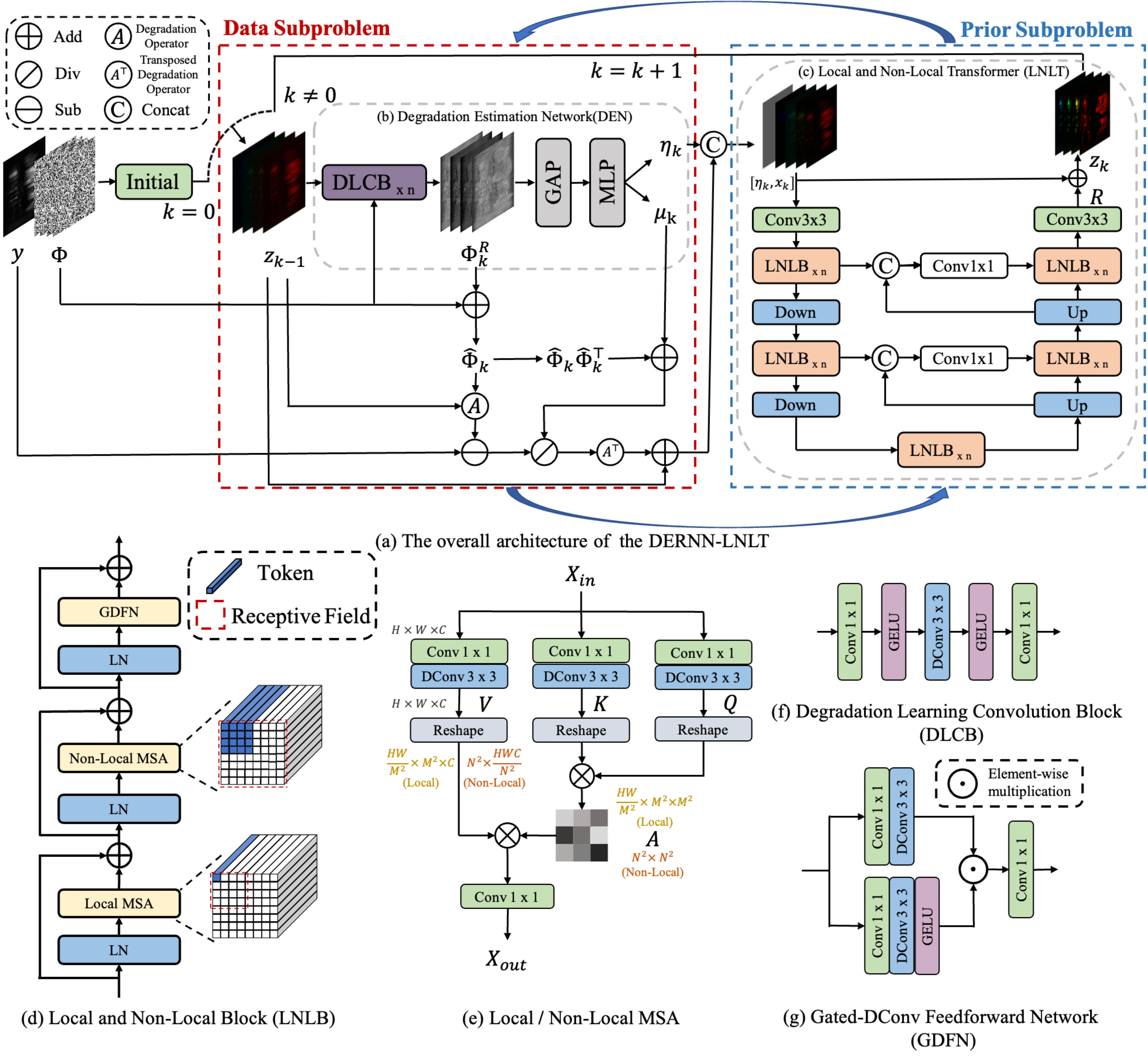}
    \caption{\small The diagram of the proposed HSI reconstruction method. (a) The overall architecture of the proposed DERNN-LNLT. (b) The diagram of the DEN. (c) The diagram of the LNLT. (d) The diagram of the LNLB. (e) The diagram of the Local/Non-Local MSA. (f) The structure of the DLCB. (g) The structure of the GDFN.}
    \label{fig:DERNN-LNLT}
\end{figure*}

In this section, we \textbf{begin} by unfolding the HQS algorithm within the MAP framework and transforming the DUN into an RNN by sharing parameters across stages. The recurrent mechanism enhances the generalization and robustness of each stage’s DNN, allowing for stable training without complex connections between stages, and dramatically reduces parameters of the DUN by several folds. \textbf{Then}, we incorporate the Degradation Estimation Network (DEN) into the RNN (DERNN), which estimates the degradation matrix for the data subproblem and the noise level for the prior subproblem by residual learning with reference to the sensing matrix. The estimated degradation matrix and noise level can improve the accuracy of solving the data subproblem and prior subproblem. \textbf{Subsequently}, we propose the Local and Non-Local Transformer (LNLT), which utilizes the Local and Non-Local Multi-head Self-Attention (MSA) to effectively exploit both local and non-local HSIs priors.  The Non-local MSA of the LNLT greatly reduces the computational complexity of the global MSA \cite{vit} in HSIs by altering the projection dimension and fixing the number of windows rather than the window size. Compared to recent local and non-local methods \cite{dauhst, maxim}, the LNLT can model non-local relationships at patch-level, ensuring that every pixel contributes to receptive fields and enabling the modeling of object-level non-local relationships. \textbf{Finally}, incorporating the LNLT into the DERNN as the denoiser for the prior subproblem leads to the proposed DERNN-LNLT.

\subsection{Degradation Estimation Recurrent Neural Network}

DUNs unfold model-based methods by stacking multiple DNNs. However, the isolation of DNNs across different stages, where the DNN at each stage struggles to receive adequate training, leads to a significant number of neurons remaining non-activated, thereby resulting in suboptimal reconstruction. Additionally, the multi-stage cascading architecture leads to unstable training and optimization \cite{mprnet}. Lastly, stacking of multiple DNNs leads to a substantial increase in the number of parameters.  We simultaneously address these issues through a simple operation: transforming DUN to an RNN by sharing the parameters of DNN across stages. This approach can be considered as multitask learning for denoising images with different noise levels. Through multitask learning, the DNN is required to have denoising capabilities for images with varying noise levels, leading to the activation of more neurons. As a result, the generalization and robustness of the DNN are improved.  \cite{dgsmp, mprnet, dgunet, rdluf_mixs2} introduce inter-stage connections or interactions for stable training, which contributes to incorporating information from other stages into the current stage. The design of RNN inherently allows DNN to process information from different stages, enabling stable training without complex connections and interactions. Moreover, RNN significantly reduces the number of required parameters and memory consumption compared to DUNs.

The overview of the DERNN is illustrated in Fig. \ref{fig:DERNN-LNLT} (a). The DERNN unfolds the HQS algorithm under the MAP theory and shares parameters across stages, alternatively solving a data problem and a prior subproblem in each recurrent step.

The original HSI signal could be estimated by minimizing the following energy function:
\begin{equation}
    \hat x = \arg \min_x \frac{1}{2} \| y - \Phi x \|^2 + \lambda R(x),
    \label{eq: energy function}
\end{equation}
where $\frac{1}{2}\|y-\Phi x\|^2$ is the data fidelity term, $R(x)$ is the prior term. In order to decouple the data term and prior term, HQS introduces an auxiliary variable $z$, resulting in a constrained optimization problem given by
\begin{equation}
    \hat x = \arg \min_{x} \frac{1}{2} \| y - \Phi x \|^2 + \lambda R(z) \quad s.t. \quad z = x.
    \label{eq: aux z}
\end{equation}
Eq. (\ref{eq: aux z}) is solved by minimizing
\begin{equation}
    \mathcal{L}_\mu(x, z) = \frac{1}{2} \| y - \Phi x\|^2 + \lambda R(z) + \frac{\mu}{2}\| z - x \|^2,
    \label{eq: penalty mu}
\end{equation}
where $\mu$ is a penalty parameter. Subsequently, Eq. (\ref{eq: penalty mu}) can be addressed by iteratively solving a data subproblem (x-subproblem) and a prior subproblem (z-subproblem):
\begin{subequations}
    \begin{gather}
        {x_{k} = \arg \min_x \|y - \Phi x\|^2 + \mu \|x - z_{k-1}\|^2,}
        \label{eq: dp} \\
        {z_k = \arg \min_z \frac{1}{2(\sqrt{\lambda / \mu})^2}} \| z - x_k \|^2 + R(z).
        \label{eq: pp}
    \end{gather}
\end{subequations}

The data subproblem, Eq. (\ref{eq: dp}), usually has a fast closed-form solution as
\begin{equation}
    x_k = (\Phi^\top \Phi + \mu I)^{-1}(\Phi^\top y + \mu z_{k-1}),
    \label{eq: close-form}
\end{equation}
where $I$ is an identity matrix with desired dimensions. By following the matrix inverse formula, the matrix inversion in Eq. (\ref{eq: close-form}) can be written as
\begin{equation}
    (\Phi^\top \Phi + \mu I)^{-1} = \mu^{-1}I - \mu^{-1}\Phi^\top(I + \Phi \mu^{-1} \Phi^\top)^{-1} \Phi \mu^{-1}.
    \label{eq: partial of close-form}
\end{equation}

\begin{figure}[!h]
    \centering
    \includegraphics[width=\linewidth]{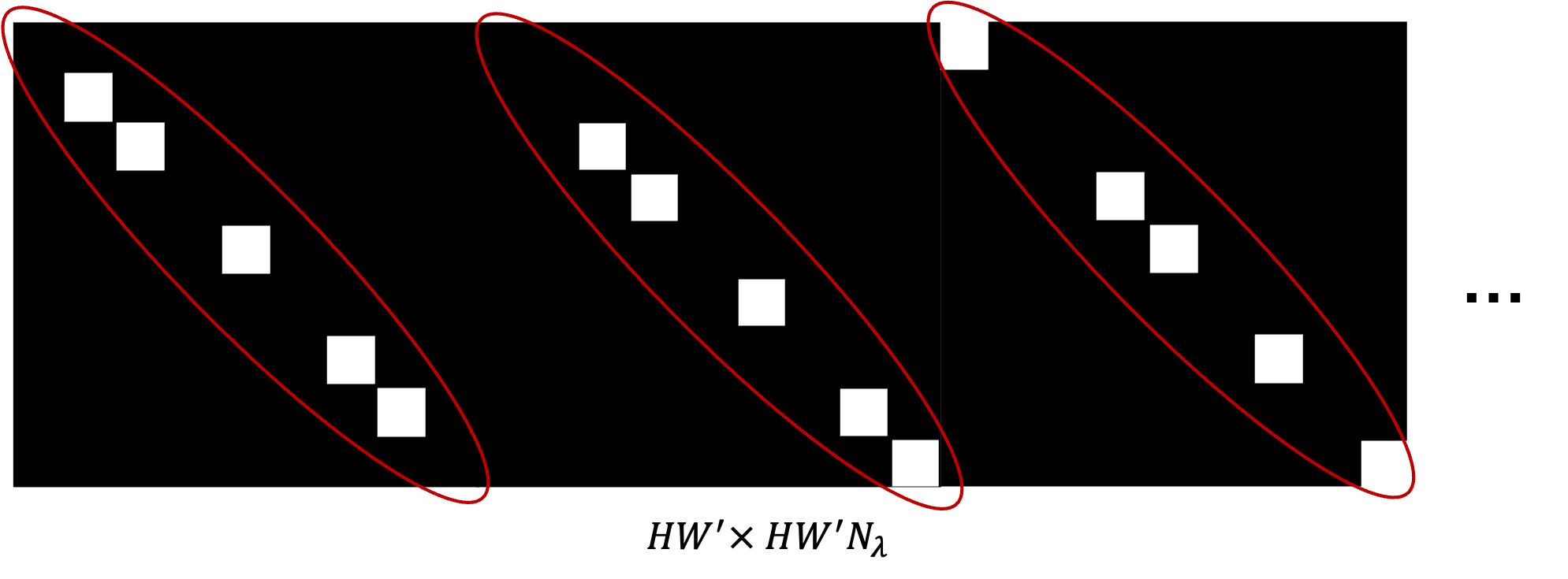}
    \caption{\small The diagram of the sensing matrix $\Phi$. }. 
    \label{fig:Phi}
\end{figure}

In CASSI systems, $\Phi \in \mathbb{R}^{HW' \times HW'N_{\lambda}}$ is a block diagonal matrix, as illustrated in Fig. \ref{fig:Phi}.  Thus, $\Phi \Phi^\top$ is a diagonal matrix that can be defined as 
\begin{equation}
    \Phi \Phi^\top \stackrel{def}{=} \text{diag}\{\phi_1, ..., \phi_i, ..., \phi_{HW'}\}.
    \label{eq: PhiPhiT}
\end{equation}
 By plugging $\Phi \Phi^\top$ into $(I + \Phi \mu^{-1} \Phi^\top)^{-1}$, we obtain:
\begin{equation}
    (I + \Phi \mu^{-1} \Phi^\top)^{-1} = \text{diag}\{\frac{\mu}{\mu + \phi_1}, ..., \frac{\mu}{\mu + \phi_i}, ..., \frac{\mu}{\mu + \phi_{HW'}}\}.
    \label{eq: eq9}
\end{equation}
By plugging Eq. (\ref{eq: partial of close-form}), Eq. (\ref{eq: PhiPhiT}) and Eq. (\ref{eq: eq9}) into Eq. (\ref{eq: close-form}) and simplifying the formula, we have
\begin{equation}
    x_k = z_{k-1} + \Phi^\top [(y - \Phi z_{k-1}) ./ (\mu + diag(\Phi \Phi^\top))],
    \label{eq: close form}
\end{equation}
where $diag$ represents the extraction of elements from its diagonal to form a vector.

The prior subproblem Eq. (\ref{eq: pp}), from a Bayesian perspective, corresponds to Gaussian denoising on $x_k$ with noise level $\sqrt{\lambda/\mu}$. To address this,  Eq. (\ref{eq: pp}) could be rewritten as follows:
\begin{equation}
    z_k = Denoiser(x_k, \sqrt{\lambda/\mu}).
    \label{eq: denoiser}
\end{equation}

DUNs iteratively solve the data subproblem (Eq. \ref{eq: close form}) and the prior subproblem (Eq. \ref{eq: denoiser}) to solve the ill-posed inverse problems. On one hand, the accuracy of solving the data subproblem is highly related to the degradation matrix. There are two common ways to acquire the degradation matrix. One way is to use the sensing matrix as the degradation matrix \cite{admm-net, gap-net, dauhst, dgsmp-pami}. The other way is to use a DNN to directly estimate the degradation matrix \cite{dgsmp, herosnet}.  However, using the sensing matrix as the degradation matrix may not accurately reflect the degradation process when capturing real scenes due to variations in exposure time, light intensity, and device errors. Additionally, directly modeling the degradation process is challenging. On the other hand, the prior subproblem is typically treated as a denoising problem of an image with a specific noise level. Therefore, the noise level is a key clue for solving the denoising problem. However, in the reconstruction task, the noise level is not pre-given. Thus, the performance of solving the data subproblem and the prior subproblem is compromised.

\textbf{Degradation Estimation Network (DEN).} To address the above issues, the DEN is proposed to simultaneously estimate the degradation matrix for the data subproblem and the noise level for the prior subproblem by residual learning with reference to the sensing matrix.  The structure of the DEN is illustrated in Fig. \ref{fig:DERNN-LNLT} (b). Instead of treating the sensing matrix as the degradation matrix or directly using a DNN to estimate the degradation matrix, the DEN consists of several degradation learning convolution blocks (DLCBs) \cite{rdluf_mixs2} to estimate the residual $\Phi_k^R$ between the sensing matrix and the degradation matrix, DLCB is detailed as Fig. \ref{fig:DERNN-LNLT} (f). To conveniently solve Eq. (\ref{eq: dp}) and Eq. (\ref{eq: pp}), we set the $\mu_k$ and $\eta_k = \frac{1}{(\sqrt{\lambda_k/\mu_k})^2}$ as learnable iteration-specific parameters.  The estimated parameters of the DEN can enhance the accuracy of solving the data subproblem and the prior subproblem. In addition, we let the residual between the sensing matrix and the degradation matrix as an intermediate feature map for predicting the noise level. The residual $\Phi_k^R$ undergoes global average pooling (GAP) and then passes through an MLP to obtain estimates of $\mu_k$ and $\eta_k$.  This shares convolutional neural networks (CNNs) in the DEN, thereby saving on the DEN's parameters. Therefore, the Eq. (\ref{eq: close form}) and Eq. (\ref{eq: pp}) can be reformulated as:
\begin{equation}
    x_k = z_{k-1} + \hat \Phi_k^\top [(y - \hat \Phi_k z_{k-1}) / (\mu_k + \hat \Phi_k \hat \Phi_k^\top)],
\end{equation}
\begin{equation}
    z_k = Denoiser(x_k, \eta_k).
    \label{eq: denoiser_hat}
\end{equation}
where $\hat \Phi_k = \Phi + \Phi_k^R$, $\Phi_k^R = \text{DLCBs}(z_{k-1}, \Phi)$.

\subsection{Local and Non-Local Transformer}

From Eq. (\ref{eq: denoiser_hat}), one can observe that a denoiser is trained to implicitly represent the prior term as a denoising problem.

It has been demonstrated that both local and non-local priors exist in HSIs \cite{wang2016adaptive, zhang2019computational}. Convolutional neural networks \cite{l-net, tsa-net, dgsmp, herosnet} and window-based local transformers \cite{swin} have been quite successful in modeling local similarities. However, they struggle with identifying non-local similarities due to their limited receptive field. The window-based global Transformer \cite{vit} have the ability to model non-local similarities across windows. However, it mainly suffers from three issues. Firstly, the implicit modeling of spatial features within a window results in a loss of critical local textures and details, thereby resulting in poor reconstruction performance. Secondly, the computational cost of self-attention in window-based global MSA increases quadratically with spatial size. Thirdly, in HSIs, where the number of channels far exceeds that of RGB images, the computational cost of the projection in window-based global MSA is also unaffordable. Recently, some works \cite{maxim, dauhst} have divided local and non-local branches by splitting attention heads, using local windows to model both local and non-local similarities by shuffling/permuting pixels. However, these methods can only model non-local similarities of position-specific pixels and model non-local similarities at the pixel-level instead of the patch-level \cite{nlm}, which may neglect some relevant non-local pixels and miss some long-range, object-level relationships, leading to poor reconstruction results.

To address the mentioned issues, we propose an efficient Local and Non-Local Transformer (LNLT). The LNLT utilizes window-based local MSA to model local similarities and global MSA to model non-local similarities.  In contrast to the window-based global MSA, which projects along the dimension of flattened elements in each window and uses a fixed window size, we alter the direction of the projection to the channel dimension and fix the number of windows rather than the window size.  This ensures that the computational cost of projection remains the same regardless of the window size and enables the computational cost of self-attention to increase linearly with the spatial size. Additionally, it allows the window of the global MSA to cover multiple windows of the local MSA, thereby better modeling non-local similarities across windows. Finally, the proposed LNLT models non-local relationships at the patch-level, ensuring that every pixel contributes to the receptive fields and enabling the modeling of object-level non-local relationships.

\textbf{Overall Architecture.} As shown in Fig. \ref{fig:DERNN-LNLT} (c), the LNLT adopts a three-level U-shaped structure, and each level consists of multiple basic units called Local and Non-Local Transformer Blocks (LNLBs). Up- and down-sampling modules are positioned between LNLBs. Firstly, the LNLT employs a $Conv3 \times 3$ to embed $x_k$ concatenated with the $\eta_k$ into the shallow feature $X_0 \in \mathbb{R}^{H \times W \times C}$. Secondly, $X_0$ passes through all the LNLBs to be embedded into the deep feature $X_d \in \mathbb{R}^{H \times W \times C}$.  Finally, a $Conv3 \times 3$ maps $X_d$ to a residual image $R \in \mathbb{R}^{H \times W \times N_{\lambda}}$. The denoised output image $z_k$ is obtained by adding $x_k$ and $R$ element-wise.

\textbf{Local and Non-Local Transformer Block.} The Local and Non-Local Transformer Block (LNLB) is the most important component, as shown in Fig. \ref{fig:DERNN-LNLT} (d). Each LNLB consists of three layer-normalizations (LNs), a Local MSA, a Non-Local MSA, and a GDFN \cite{restormer}. Previous works \cite{mst, restormer} have shown that the directions of projection and self-attention can be different. For instance, the projection can be along the channel dimension while the inner product can be performed along the flattened spatial dimension. Thus, the Local MSA and the Non-Local MSA both employ the same way to embed inputs into queries, keys, and values.  They project the input along the channel dimension, which significantly reduces the computational cost of the projection of Non-local MSA  in HSIs, ensuring that the computational cost of projection remains the same for any window size. The difference lies in how they compute the attention maps. Fig. \ref{fig:DERNN-LNLT} (e) illustrates the computation process of the Local MSA and the Non-Local MSA. Let $X_{in}\in\mathbb{R}^{H\times W\times C}$ denotes the input of the Local/Non-Local MSA. The Local/Non-Local MSA utilizes a $Conv 1 \times 1$ and a $DConv 3 \times 3$ to embed $X_{in}$ into query $\mathrm{Q}\in\mathbb{R}^{H\times W\times C}$, key $\mathrm{K}\in\mathbb{R}^{H\times W\times C}$, and value $\mathrm{V}\in\mathbb{R}^{H\times W\times C}$, where $DConv$ represents the depth-wise convolution \cite{mobilenets}:
\begin{equation}
    \mathrm{Q} = W_d^\mathrm{Q}W_p^\mathrm{Q}X_{in}, \mathrm{K} = W_d^\mathrm{K}W_p^\mathrm{K}X_{in},  \mathrm{V} = W_d^\mathrm{V}W_p^\mathrm{V}X_{in},
\end{equation}
where $W_p^{(\cdot)}$ is the $Conv 1 \times 1$ and $W_d^{(\cdot)}$ is the $DConv 3 \times 3$.

\textbf{Local Multi-head Self-Attention (Local MSA).} The Local MSA partitions the input into non-overlapping windows of size $M \times M$, treating each pixel within the window as a token, and computes self-attention within the window. In the Local MSA, the query $\mathrm{Q}$, key $\mathrm{K}$, and value $\mathrm{V}$ are reshaped as $\mathrm{Q_L}, \mathrm{K_L}, \mathrm{V_L} \in \mathbb{R}^{\frac{HW}{M^2} \times M^2 \times C}$. Subsequently, $\mathrm{Q_L}$, $\mathrm{K_L}$, and $\mathrm{V_L}$ are split along the last dimension into $h$ heads: $\mathrm{Q_L} = [\mathrm{Q_L^1}, ..., \mathrm{Q_L^h}]$, $\mathrm{K_L} = [\mathrm{K_L^1}, ..., \mathrm{K_L^h}]$, $\mathrm{V_L} = [\mathrm{V_L^1}, ..., \mathrm{V_L^h}]$. The dimension of each head is  $\mathrm{d_L^h} = \frac{C}{h}$. The inner product of the query $\mathrm{Q_L^i}$ and key $\mathrm{K_L^i}$ generates an attention map $\mathrm{A_L^i}$ with a shape of $\mathbb{R}^{\frac{HW}{M^2}\times M^2\times M^2}$, which represents the correlations between different pixels within the window. Overall, the Local MSA process is defined as:
\begin{equation}
    \begin{aligned}
        X_{out} &= W_p \text{Concat}(\text{head}_1, ..., \text{head}_n) + X_{in}, \\
        \text{head}_i &= \text{Attention}(\mathrm{Q_L^i}, \mathrm{K_L^i}, \mathrm{V_L^i}) \\ 
        &= \text{Softmax}(\frac{\mathrm{Q_L^i} \mathrm{K_L^i}}{\sqrt{\mathrm{d_L^h
}}} + \mathrm{P^i_L})\mathrm{V_L^i},
    \end{aligned}
\end{equation}
where $\mathrm{P_L^i} \in \mathbb{R}^{M^2 \times M^2}$ are learnable position embedding, representing the positional relationships between different pixels within the window.  

\textbf{Non-local Multi-head Self-Attention (Non-Local MSA).} The Non-Local MSA divides the input into $N \times N$ non-overlapping windows, treating each window as a token, and computes self-attention across the windows. In the Non-Local MSA,  the query $\mathrm{Q}$, key $\mathrm{K}$, and value $\mathrm{V}$ are reshaped as $\mathrm{Q_{NL}}, \mathrm{K_{NL}}, \mathrm{V_{NL}} \in \mathbb{R}^{N^2 \times \frac{HWC}{N^2}}$. Subsequently, $\mathrm{Q_{NL}}$, $\mathrm{K_{NL}}$, and $\mathrm{V_{NL}}$ are split along the last dimension into $h$ heads: $\mathrm{Q_{NL}} = [\mathrm{Q_{NL}^1}, ..., \mathrm{Q_{NL}^h}]$, $\mathrm{K_{NL}} = [\mathrm{K_{NL}^1}, ..., \mathrm{K_{NL}^h}]$, $\mathrm{V_{NL}} = [\mathrm{V_{NL}^1}, ..., \mathrm{V_{NL}^h}]$. The dimension of each head is  $\mathrm{d_{NL}^h} = \frac{HWC}{hN^2}$. The inner product of the query $\mathrm{Q_{NL}^i}$ and key $\mathrm{K_{NL}^i}$ generates an attention map $\mathrm{A_{NL}^i}$ with a shape of $\mathbb{R}^{N^2\times N^2}$, which represents the correlations between different windows. Overall, the Non-Local MSA process is defined as:
\begin{equation}
    \begin{aligned}
        X_{out} &= W_p \text{Concat}(\text{head}_1, ..., \text{head}_n) + X_{in}, \\
        \text{head}_i &= \text{Attention}(\mathrm{Q_{NL}^i}, \mathrm{K_{NL}^i}, \mathrm{V_{NL}^i}) \\ 
        &= \text{Softmax}(\frac{\mathrm{Q_{NL}^i} \mathrm{K_{NL}^i}}{\sqrt{\mathrm{d_{NL}^h}}} + \mathrm{P^i_{NL}})\mathrm{V_{NL}^i},
    \end{aligned}
\end{equation}
where $\mathrm{P_{NL}^i} \in \mathbb{R}^{N^2 \times N^2}$ are learnable position embedding, representing the positional relationships between different windows. 

\section{Experimental Results}

\subsection{Experimental Setting}

\textbf{Simulation Setting.} In the simulation experiments, the compared models were all trained on the CAVE \cite{cave} dataset and subsequently tested on the KAIST \cite{kaist} dataset. For fair comparisons, we selected 10 scenes of a spatial size $256 \times 256$ from the KAIST dataset for testing, following previous works \cite{tsa-net, dgsmp, mst, rdluf_mixs2}. The windows size for the Local MSA and the number of windows for the Non-Local MSA were set $M = N = 8$. To evaluate the quality of the reconstructed images, we employed the metrics of PSNR, SSIM \cite{ssim} and SAM. 

\textbf{Real Setting.} In the real experiments, the compared models were all trained on CAVE \cite{cave} and KAIST \cite{kaist} datasets. For testing, we used 5 real measurements with a spatial size of $660 \times 714$, captured by a real CASSI system \cite{tsa-net}. To simulate real measurement conditions, 11-bit shot noise was injected in the training measurements.  The windows size for the Local MSA and the number of windows for the Non-Local MSA were set $M = N = 14$.

\textbf{Data Preprocessing.} During the training process, we employed random cropping of 3D HSI datasets to generate patches of dimensions $256 \times 256 \times 28$ and $660 \times 660 \times 28$, which served as labels for the simulation and real experiments, respectively. Correspondingly, real shifted masks \cite{tsa-net} of dimensions $256 \times 310 \times 28$ and $660 \times 714 \times 28$ were utilized for simulation and real experiments. The dispersion shift steps were set to 2. We use data augmentation techniques, including random flipping, random rotation, and mosaic augmentation \cite{yolov4} were employed following previous works \cite{mst, cst, dauhst, rdluf_mixs2}.

\textbf{Optimization.} The objective of the model was to minimize the Charbonnier loss. We employed the Adam optimizer with hyperparameters $\beta_1 = 0.9$ and $\beta_2 = 0.999$. The training process spanned 300 epochs in total. The cosine annealing scheduler with linear warm-up was utilized.

\textbf{Hardware and Software.} For hardware, we trained our model on a Tesla A100 80G GPU with an AMD EPYC 7763 64-Core CPU and 1.6TB of memory. For software, the operating system is Ubuntu 20.04 and we used Python 3.8 and PyTorch 1.12.1. 

\subsection{Quantitative Results}

To demonstrate the effectiveness of the proposed method, we conducted comparisons with nine existing approaches on simulation datasets.  These include three model-based methods (TwIST \cite{twist}, GAP-TV \cite{gap-tv}, DeSCI \cite{desci}), two end-to-end networks methods (TSA-Net \cite{tsa-net} and MST \cite{mst}), and four deep unfolding networks (GAP-Net \cite{gap-net}, ADMM-Net \cite{admm-net}, DAUHST \cite{dauhst}, RDLUF-Mix$S^2$ \cite{rdluf_mixs2}). The corresponding results for 10 simulated scenes are presented in Table \ref{Tab:performance}. It is noteworthy that while both end-to-end networks and DUNs exhibited superior performance compared to model-based methods, the proposed method with 7 stages surpassed them all with fewer GFlops and Params. Specifically, compared to TSA-Net \cite{tsa-net}, GAP-Net \cite{gap-net}, ADMM-Net \cite{admm-net}, MST-L \cite{mst}, DAUHST-9stg \cite{dauhst}, and RDLUF-Mix$S^2$ 9stg, the DERNN-LNLT 9stg$^*$ outperformed them with improvements of 8.87 dB, 7.07 dB, 6.75 dB, 5.15 dB, 1.97 dB, and 0.76 dB on average, respectively. Additionally, the proposed method achieves SOTA performance in PSNR, SSIM and SAM while significantly reducing the number of parameters, as shown in Fig. \ref{fig:teaser} and Table. \ref{Tab:performance}, which reduces the parameters by several folds compared to other methods.

\input{tables/table1}

\subsection{Qualitative Results}

\begin{figure*}[!htb]
    \centering
    \includegraphics[width=\linewidth, height=0.55\linewidth]{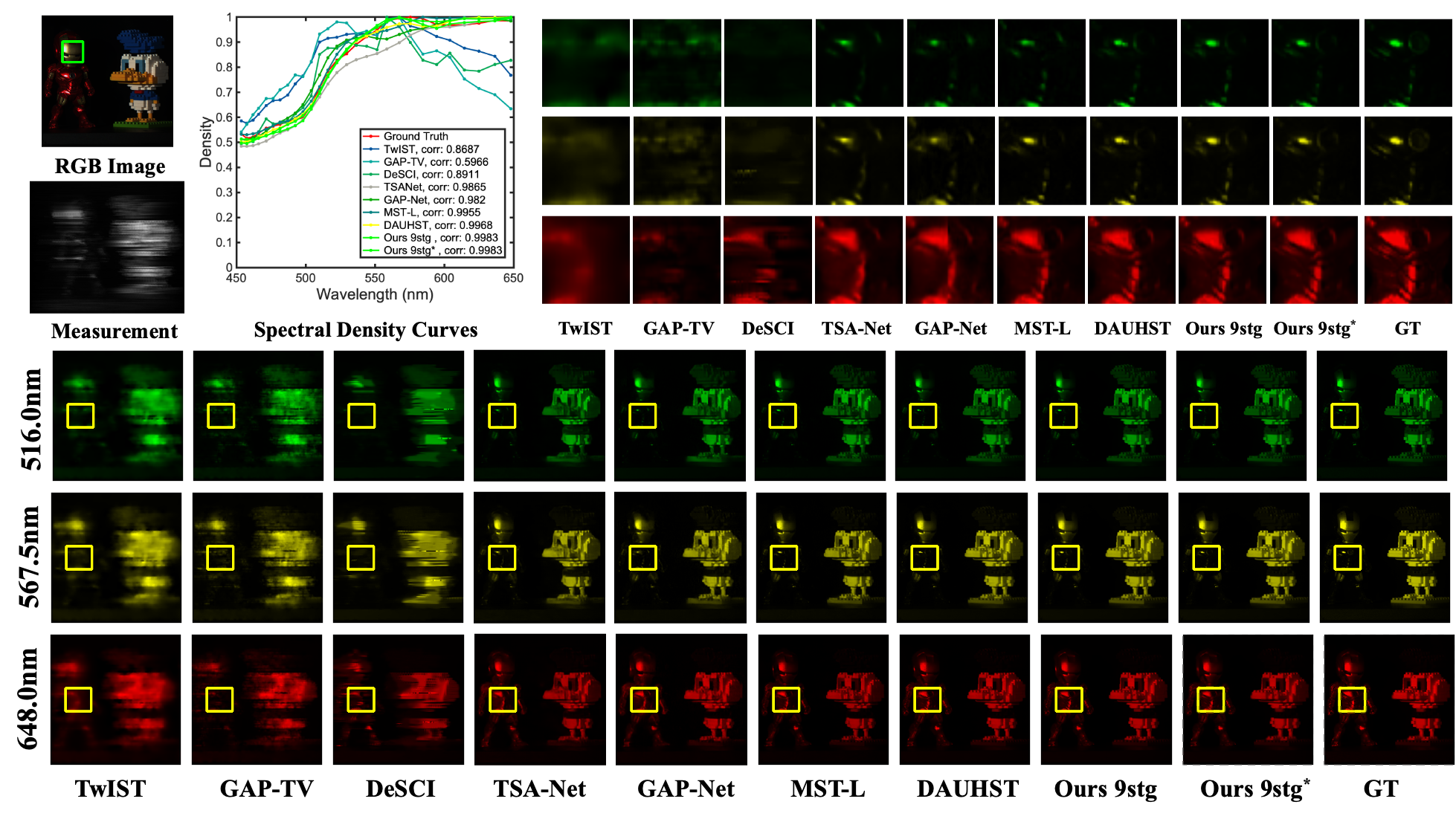}
    \caption{\small Comparisons of reconstructed HSIs use 3 out of 28 spectral channels in Scene 8. The top-middle shows the spectral curves corresponding to the green boxes of the RGB image. The top-right depicts the enlarged patches corresponding to the yellow boxes in the bottom HSIs. Zoom in for a better view.}
    \label{fig:simulation result}
\end{figure*}

\textbf{Simulation Results}. In Fig. \ref{fig:simulation result}, we present a comparison of the proposed method using 3 out of 28 spectral channels from Scene 8 with the simulation results obtained from seven SOTA approaches. The top-right of the figure shows zoomed-in patches of the yellow boxes within the entire HSIs (bottom). As can be observed that the proposed method effectively produces perceptually pleasing images with more vivid and sharp edge details, while maintaining spatial smoothness in homogeneous regions without introducing artifacts.  This is attributed not only to the design of the RNN, which enhances the representativeness of the DUN, but also to the exploitation of local and non-local priors by the LNLT. In contrast, previous methods either introduce undesired chromatic artifacts and blotchy textures that are absent in the ground truth or yield over-smooth results, compromising fine-grained structures. The top-middle part illustrates the spectral density curves corresponding to the green boxes in the RGB image (top-left). The spectral curves of the proposed method exhibit the highest correlation and coincidence with the reference curves. This can be attributed to the estimated parameters by the degradation estimation network (DEN), which leads to more accurate solutions for both the data subproblem and the prior subproblem, further demonstrating the effectiveness of the degradation estimation strategy.

\begin{figure*}[!htb]
    \centering
    \includegraphics[width=\linewidth, height=0.388\linewidth]{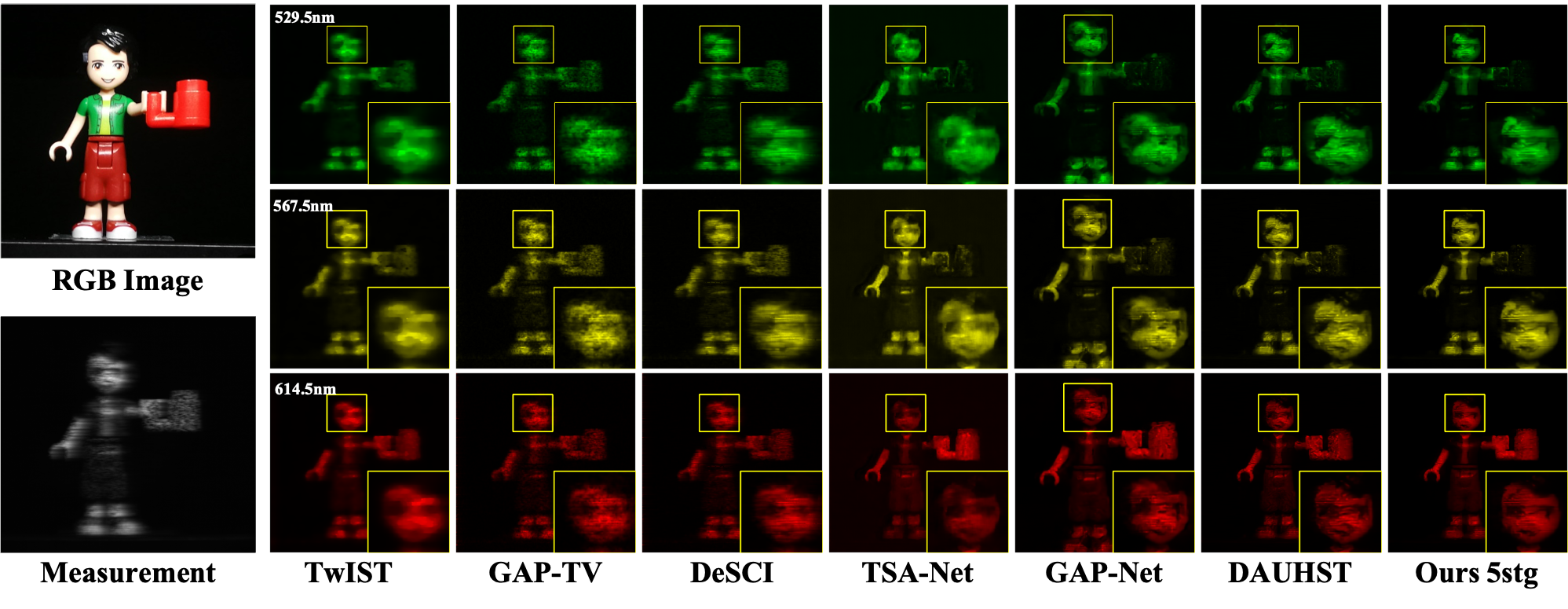}
    \caption{\small Real HSI reconstruction comparison of Scene 3. 3 out of 28 spectra are randomly selected. }
    \label{fig:real result}
\end{figure*}

\textbf{Real Results}. To evaluate the effectiveness of the proposed method on real data, we compared the reconstructed images of the real scene (3 of 28 spectral channels of Scene 3) using our DERNN-LNLT and six SOTA approaches. As shown in Fig. \ref{fig:real result}, our results can reconstruct clearer contents and detailed textures with fewer artifacts and blurring, which benefits from the exploration of local and non-local features. Specifically, our approach restores a clearer right eye and a more natural mouth shape with fewer distortion. Our robust real dataset results are attributed to the DERNN learning features from different stages, enhancing the generalization and robustness of the DUN, and the estimated parameters of DEN narrowing the gap between the sensing matrix and the degradation process, thereby improving the accuracy of solving the data sub-problem, and the explicit noise level enhancing the accuracy of the prior sub-problem.

\begin{figure*}[!htb]
    \centering
    \includegraphics[width=\linewidth, height=0.5\linewidth]{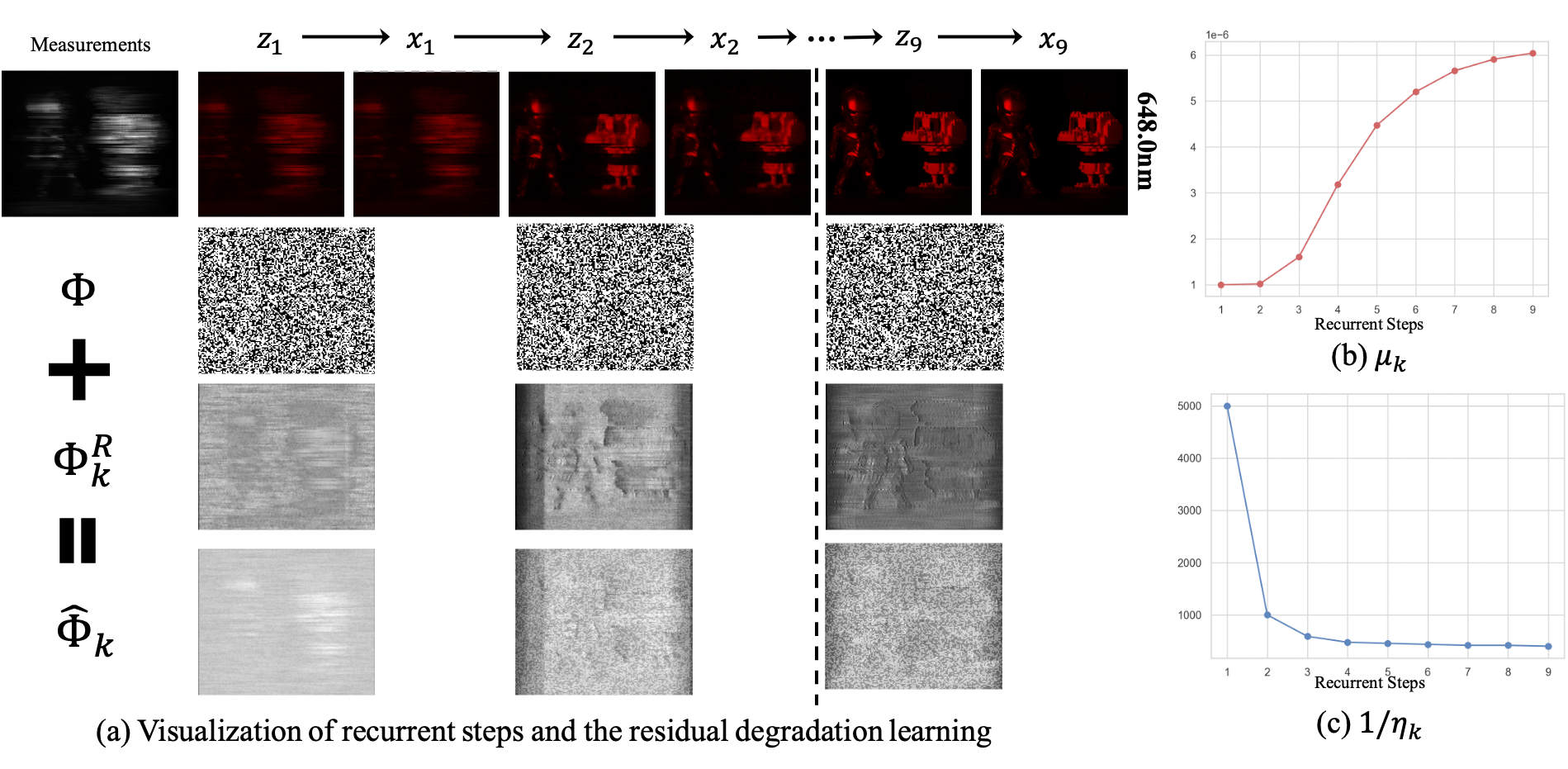}
    \caption{\small The visualization of the estimated parameters of the Degradation Estimation Network. (a) The results of the reconstruction, the visualizations of the sensing matrix $\Phi$, the estimated residual $\Phi_k^R$, and the estimated degradation matrix $\hat \Phi_k$. (b) The values of $\mu_k$ at k-th recurrent steps. (c) The values of $1 / \eta_k$ at k-th recurrent steps.}
    \label{fig:visualization of the DE}
\end{figure*}

\begin{figure*}[!htb]
    \centering
    \includegraphics[width=\linewidth, height=0.568\linewidth]{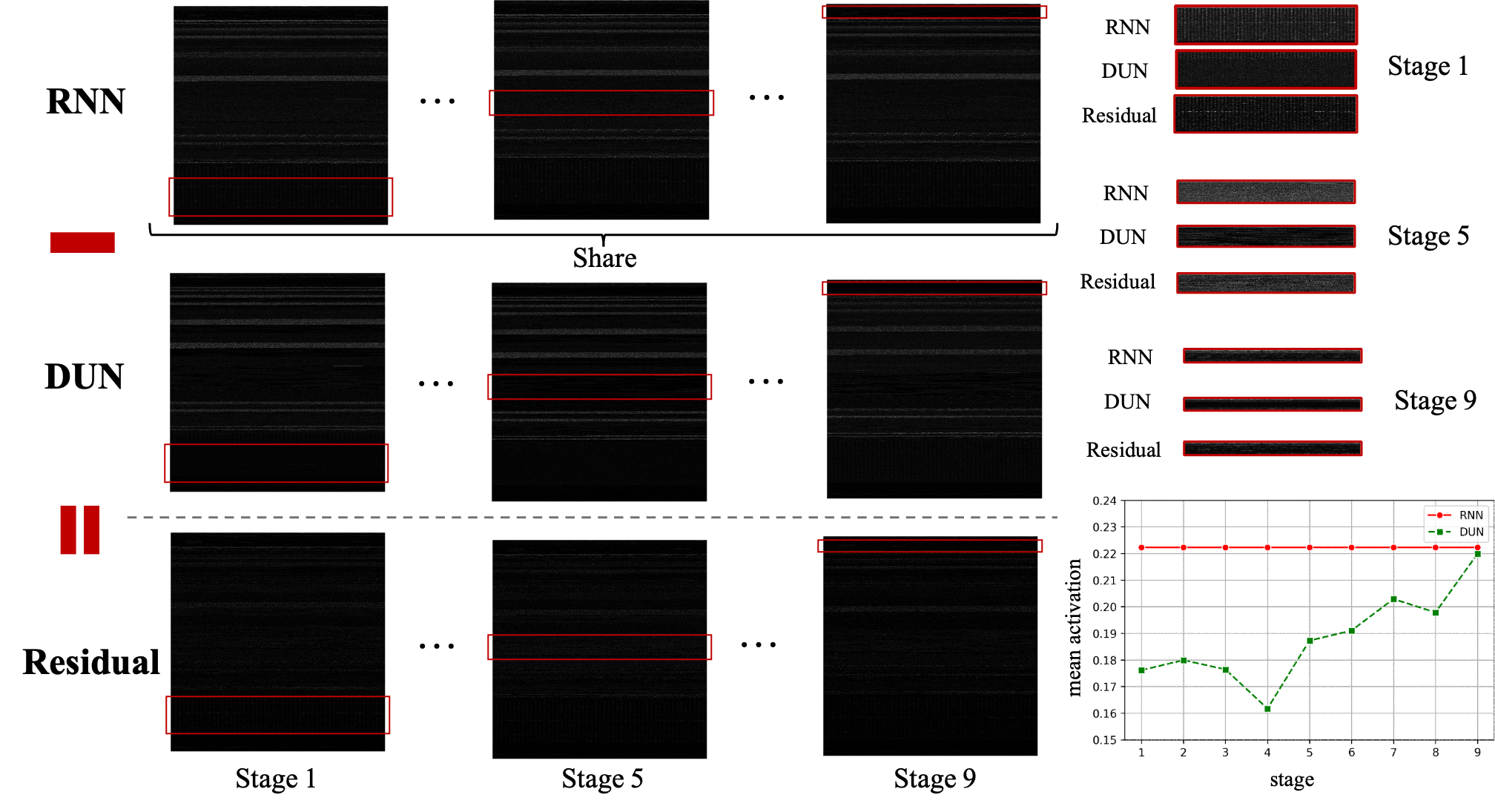}
    \caption{\small Visualization of parameters, RNN, DUN, and their Residual are demonstrated. The top right corner shows the part of parameters in the red box in the left, scaled by a factor. The bottom right corner shows the mean activation of parameters for RNN and DUN respectively. Please zoom in and increase the screen brightness for a better view.}
    \label{fig:Visualization of parameters}
\end{figure*}

\begin{figure*}[!htb]
    \centering
    \includegraphics[width=\linewidth, height=0.32\linewidth]{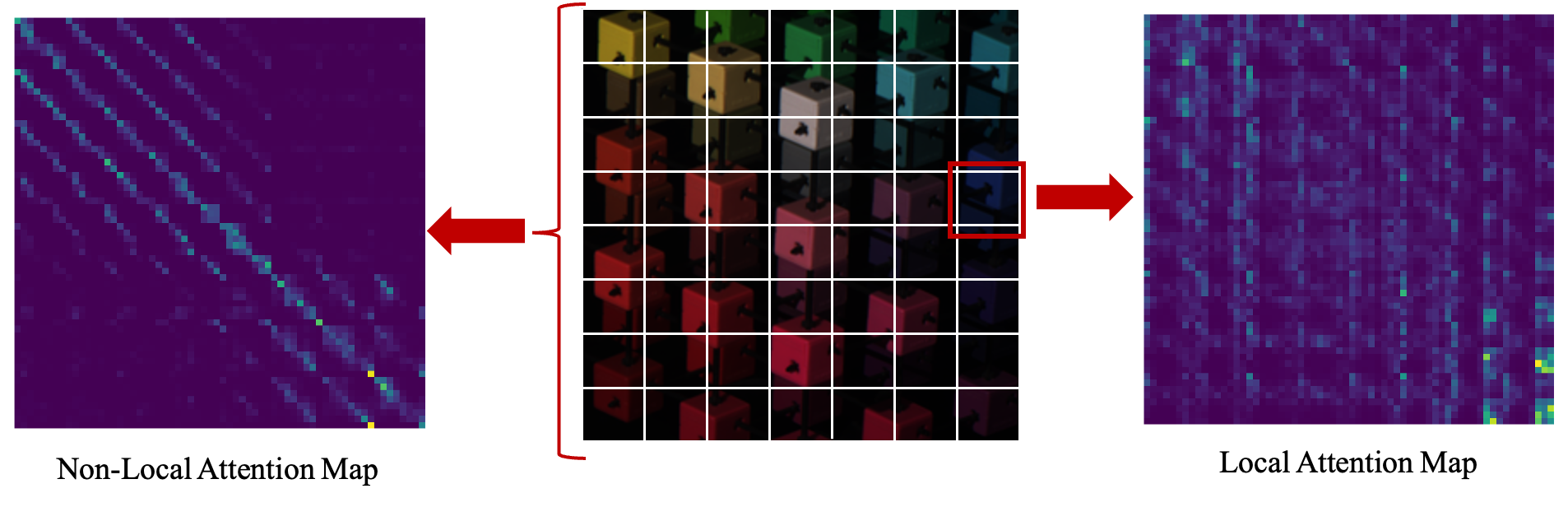}
    \caption{\small The visualization of the Local and Non-local attention map on the simulation scene 2.}
    \label{fig:Attention map}
\end{figure*}

\subsection{Analysis of DERNN} 

In this section, we first explain the intuition behind the estimated penalty parameter $\mu_k$ in the Degradation Estimation Network (DEN).  Then, we analyze the DEN by visualizing the sensing matrix, the estimated residual and the degradation matrix, and by plotting the curves of the noise level and penalty parameter as they change with the recurrent steps. Finally, we analyze the effectiveness of the Recurrent Mechanism by visualizing the parameters of the RNN and DUN, such as activating more neurons (parameters).  

In the DEN, we also set the penalty parameter $\mu$ as learnable iteration-specific parameters $\mu_k$, because manually tweaking it proves to be non-trivial. To ensure that $x_k$ and $z_k$ converge to a fixed point, a large value of $\mu$ is required, but this could lead to a higher number of iterations $(K)$ for convergence \cite{DPIR, dauhst}. A common approach is to use the continuation strategy, gradually increasing $\mu$ to create a sequence of $\mu_1 < \mu_2 < \ldots < \mu_k < \ldots < \mu_K$. However, this strategy introduces an additional parameter to regulate the step size, making the parameter setting more complex. To address this issue, instead of manually tweaking the penalty parameter $\mu$, the DEN also estimates it as learnable iteration-specific parameters by residual learning from the input of the current recurrent step and with reference to the sensing matrix.  As shown in Fig. \ref{fig:visualization of the DE} (b), the penalty parameter $\mu$ is gradually increased, forcing $x_k$ and $z_k$ to converge to a fixed point. This tendency is consistent with the manual continuation strategy, indicating that the proposed estimation method can automatically adjust the magnitude of $\mu$.

To analyze the DEN, in Fig.  \ref{fig:visualization of the DE} (a), we visualized the sensing matrix, the estimated residual, and the degradation matrix. We also plot the curves of $\mu$ and $1/\eta$ as they change with the recurrent steps in Fig. \ref{fig:visualization of the DE} (b) and Fig. \ref{fig:visualization of the DE} (c).  From the bottom part of Fig. \ref{fig:visualization of the DE} (a), it can be observed that the estimated residual captures key clues for HSIs, such as foreground and background. The corrected degradation matrix by the residual makes the degradation operator concentrate more on informative regions during each step of solving the data subproblem, resulting in improved accuracy of the solution. As shown in Fig. \ref{fig:visualization of the DE} (c), the noise level is very high in the first recurrent step. With more recurrent steps, the DERNN with the powerful LNLT denoiser, effectively reduces the noise level in the HSI.  These results demonstrate that $\eta$ can provide key information about the noise level for the denoising problem.

To analyze the Recurrent Mechanism, we visualized the intermediate state of the $x_k$ and $z_k$ of the DUN and RNN, as well as the parameters of different stages of the RNN and DUN, as shown in Fig. \ref{fig:visualization of the DE} and Fig. \ref{fig:Visualization of parameters}. From the recurrent steps on the top part of Figure \ref{fig:visualization of the DE} (a), we can observe that the RNN has achieved a good reconstruction result even after only one recurrent step. This is attributed to the RNN's ability to learn reconstruction from the input across different stages, with more neurons being activated, thereby enhancing its representational capacity and generalization. The left side of Fig. \ref{fig:Visualization of parameters} demonstrates the visualization of parameters at different stages of the DNN and RNN. The top right corner of Fig. \ref{fig:Visualization of parameters} displays an zoom in and brightened view of the parameters in the red-box area from the left side's visualization. The bottom right corner of Fig. \ref{fig:Visualization of parameters} illustrates the mean activation of parameters at each stage of the DUN and RNN. Overall,  part of parameters are activated in both the DUN and RNN, indicating that these parameters play a key role in the reconstruction.  However, the DNNs at different stages of the DUN activate fewer parameters compared to the RNN. Specifically, within the red-box area, the RNN shows more vertical bar-shaped parameters at stage 1, more scattered parameters at stage 5, and more horizontal line-shaped parameters at stage 9. This can also be observed from the bottom right corner of Fig. \ref{fig:Visualization of parameters},  the mean activation of the parameters at each stage of the DUN are consistently lower than those of the RNN.

\subsection{Visualization of Local and Non-Local MSA} 

To analyze the roles of the Local and Non-Local MSA in the LNLT, we visualize the attention maps of the Local MSA and the Non-Local MSA, as shown in Fig. \ref{fig:Attention map}. We visualized the Local and Non-Local MSA at the third level of the Local and Non-Local Transformer (LNLT). This is because for an input of size $256 \times 256$, at the third level, the feature map was downsampled two times to become $64 \times 64$, at which point the window size of the Non-Local MSA is equal to that of the Local MSA when $N = M = 8$. From the left part of Fig. \ref{fig:Attention map}, it can be observed that the Non-Local MSA aggregates information from other patches, exhibiting multiple diagonal patterns. This characteristic signifies its capability to model not only self-similarity but also the similarity between different patches, enabling thorough modeling of non-local similarities. From the right part of Fig. \ref{fig:Attention map}, we can observe that important information is aggregated onto specific tokens, with attention occurrences that are not dependent on queries, resulting in vertical patterns in the attention map. This suggests that certain pixels are essential for a patch and are not affected by other pixels.

\subsection{Ablation Study}

To investigate the specific impact of the different components of the DERNN-LNLT on its overall performance, we conducted an ablation study, and the detailed results are presented in Table \ref{tab:Break-down ablation}. 

\input{tables/table2}

\textbf{Degradation Estimation Network}. To assess the impact of the DEN, we gradually incorporate the estimated parameter into the aforementioned HQS-LNLT, leading to a Degradation Estimation Unfolding Framework (DEUF-LNLT). Initially, we incorporate degradation learning into the HQS-LNLT, estimating the residual between the sensing matrix and the degradation matrix, leading to a 0.37 dB improvement. Subsequently, the integration of the estimated penalty parameter yields an additional improvement of 0.07 dB. Additionally, the incorporation of the estimated noise level contributes to a 0.08 dB improvement.  Ultimately, the DEUF-LNLT showcases a 0.52 dB improvement compared to the HQS-LNLT, showing the effectiveness of the proposed DEN.

\textbf{The efficacy of the LNLT.} To validate the effectiveness of Non-Local MSA, we replaced the Non-Local MSA in LNLB with Local MSA, resulting in (2) in Tab. \ref{tab: Ablation}. The LNLB is shown in (3) in Tab. \ref{tab: Ablation}. The Non-Local MSA achieved better performance with the same memory and computation cost. Additionally, we replaced Local and Non-Local MSA in LNLB with other local and non-local priors like HS-MSA \cite{dauhst} and multi-axis gMLPs \cite{maxim}, as shown in (4) and (5) in Tab. \ref{tab: Ablation}. LNLT achieved superior performance with fewer memory and computation costs. 

\input{tables/table5}

\textbf{Recurrent Mechanism (RM)}. To assess the impact of the RM, we convert the aforementioned DEUF-LNLT into an RNN by sharing parameters across stages, resulting in the DERNN-LNLT. As presented in Table. \ref{tab:Break-down ablation}, the RM yields a performance improvement of 0.35dB and dramatically reduces the number of parameters to 1/5 of that in DEUF-LNLT. Additionally, we have also explored the benefits of varying numbers of stages. As shown in Table. \ref{tab: RM},  the performance of the network improves with an increase in the number of stages, indicating the efficacy of the RM. Lastly, we observed that a 9-stage DEUF-LNLT not only leads to a ninefold increase in parameters but also results in a significant performance decrease of 0.70dB.

\input{tables/table3}

\newpage

\section{Conclusions}

In this paper, we propose an accurate and lightweight DUN for compressive spectral imaging, named DERNN-LNLT. The DERNN-LNLT introduces a Degradation Estimation Network (DEN) to correct the gaps between the imaging model used in DUNs and the real CASSI imaging process, improving the accuracy of the data subproblem and the prior subproblem of DUNs. Additionally, the DERNN-LNLT incorporates an efficient Local and Non-local Transformer (LNLT) to solve the prior subproblem, which leverages the advantages of window-based local MSA and global MSA. The LNLT not only effectively models local and non-local similarities but also reduces the computational cost of the window-based global MSA. Furthermore, the DERNN-LNLT transforms the DUN into a Recurrent Neural Network (RNN) by sharing parameters of DNNs across stages. The design of RNN not only allows DNN to be trained more adequately but also dramatically reduces the parameters of DUNs by several folds. Experimental results on both simulation and real datasets demonstrate the superiority of the proposed DERNN-LNLT.

\bibliographystyle{IEEEtran}
\bibliography{references}

\end{document}

%% file: tables/table1.tex
\begin{table*}[t]
	\renewcommand{\arraystretch}{1.5}
	\newcommand{\tabincell}[2]{\begin{tabular}{@{}#1@{}}#2\end{tabular}}
	\centering
        \caption{Comparisons between DERNN-LNLT and SOTA methods are conducted on 10 simulation scenes. The reported results include Params, GFlops, PSNR (upper entry in each cell), SSIM (middle entry in each cell) and SAM (lower entry in each cell). * denotes each level of LNLT stacks 2 LNLBs.}
        \resizebox{\textwidth}{!}
	{
        \centering
        \begin{tabular}{c|c|c|c|c|c|c|c|c|c|c|c|c|c}
            \toprule[0.2em]
            \rowcolor{lightgray}
            ~~~~~Algorithms~~~~~
            & ~~~~~Params~~~~~
            & ~~~~~GFlops~~~~~
            & ~~~~~S1~~~~~
            & ~~~~~S2~~~~~
            & ~~~~~S3~~~~~
            & ~~~~~S4~~~~~
            & ~~~~~S5~~~~~
            & ~~~~~S6~~~~~
            & ~~~~~S7~~~~~
            & ~~~~~S8~~~~~
            & ~~~~~S9~~~~~
            & ~~~~~S10~~~~~
            & ~~~~Avg~~~~
            \\
            \midrule
            TwIST 
            & -
            & -
            &\tabincell{c}{25.16 \\ 0.700 \\ 17.145}
            &\tabincell{c}{23.02 \\ 0.604 \\ 20.011}
            &\tabincell{c}{21.40 \\ 0.711 \\ 21.005}
            &\tabincell{c}{30.19 \\ 0.851 \\ 17.481}
            &\tabincell{c}{21.41 \\ 0.635 \\ 18.611}
            &\tabincell{c}{20.95 \\ 0.644 \\ 23.561} 
            &\tabincell{c}{22.20 \\ 0.643 \\ 19.083}
            &\tabincell{c}{21.82 \\ 0.650 \\ 23.474}
            &\tabincell{c}{22.42 \\ 0.690 \\ 19.167}
            &\tabincell{c}{22.67 \\ 0.569 \\ 23.313}
            &\tabincell{c}{23.12 \\ 0.669 \\ 20.285}
            \\
            \midrule
            GAP-TV 
            & -
            & -
            &\tabincell{c}{26.82 \\ 0.754 \\ 12.874}
            &\tabincell{c}{22.89 \\ 0.610 \\ 22.250}
            &\tabincell{c}{26.31 \\ 0.802 \\ 13.793}
            &\tabincell{c}{30.65 \\ 0.852 \\ 16.087}
            &\tabincell{c}{23.64 \\ 0.703 \\ 13.465}
            &\tabincell{c}{21.85 \\ 0.663 \\ 22.741}
            &\tabincell{c}{23.76 \\ 0.688 \\ 15.589}
            &\tabincell{c}{21.98 \\ 0.655 \\ 24.528}
            &\tabincell{c}{22.63 \\ 0.682 \\ 19.301}
            &\tabincell{c}{23.10 \\ 0.584 \\ 22.021}
            &\tabincell{c}{24.36 \\ 0.669 \\ 18.265}
            \\
            \midrule
            DeSCI
            & -
            & -
            &\tabincell{c}{27.13 \\ 0.748 \\ 13.890}
            &\tabincell{c}{23.04 \\ 0.620 \\ 21.180}
            &\tabincell{c}{26.62 \\ 0.818 \\ 13.172}
            &\tabincell{c}{34.96 \\ 0.897 \\ 13.924}
            &\tabincell{c}{23.94 \\ 0.706 \\ 13.972}
            &\tabincell{c}{22.38 \\ 0.683 \\ 21.423}
            &\tabincell{c}{24.45 \\ 0.743 \\ 13.133}
            &\tabincell{c}{22.03 \\ 0.673 \\ 23.394}
            &\tabincell{c}{24.56 \\ 0.732 \\ 16.015}
            &\tabincell{c}{23.59 \\ 0.587 \\ 20.978}
            &\tabincell{c}{25.27 \\ 0.721 \\ 17.108}
            \\
            \midrule
            TSA-Net 
            & 44.25M
            & 110.06
            &\tabincell{c}{32.03 \\ 0.892 \\ 9.250}
            &\tabincell{c}{31.00 \\ 0.858 \\ 11.464}
            &\tabincell{c}{32.25 \\ 0.915 \\ 8.198}
            &\tabincell{c}{39.19 \\ 0.953 \\ 13.327}
            &\tabincell{c}{29.39 \\ 0.884 \\ 7.868}
            &\tabincell{c}{31.44 \\ 0.908 \\ 12.567}
            &\tabincell{c}{30.32 \\ 0.878 \\ 7.854}
            &\tabincell{c}{29.35 \\ 0.888 \\ 13.839}
            &\tabincell{c}{30.01 \\ 0.890 \\ 9.224}
            &\tabincell{c}{29.59 \\ 0.874 \\ 12.783}
            &\tabincell{c}{31.46 \\ 0.894 \\ 10.637}
            \\
            \midrule
            GAP-Net 
            & 4.27M
            & 78.58
            &\tabincell{c}{33.74 \\ 0.911 \\ 9.538}
            &\tabincell{c}{33.26 \\ 0.900 \\ 13.679}
            &\tabincell{c}{34.28 \\ 0.929 \\ 9.069}
            &\tabincell{c}{41.03 \\ 0.967 \\ 13.987}
            &\tabincell{c}{31.44 \\ 0.919 \\ 8.865}
            &\tabincell{c}{32.40 \\ 0.925 \\ 14.854}
            &\tabincell{c}{32.27 \\ 0.902 \\ 8.561}
            &\tabincell{c}{30.46 \\ 0.905 \\ 17.178}
            &\tabincell{c}{33.51 \\ 0.915 \\ 9.862}
            &\tabincell{c}{30.24 \\ 0.895 \\ 15.369}
            &\tabincell{c}{33.26 \\ 0.917 \\ 12.096}
            \\
            \midrule
            ADMM-Net
            & 4.27M
            & 78.58
            &\tabincell{c}{34.12 \\ 0.918 \\ 9.061}
            &\tabincell{c}{33.62 \\ 0.902 \\ 13.708}
            &\tabincell{c}{35.04 \\ 0.931 \\ 8.630}
            &\tabincell{c}{41.15 \\ 0.966 \\ 13.706}
            &\tabincell{c}{31.82 \\ 0.922 \\ 8.914}
            &\tabincell{c}{32.54 \\ 0.924 \\ 14.930}
            &\tabincell{c}{32.42 \\ 0.896 \\ 8.488}
            &\tabincell{c}{30.74 \\ 0.907 \\ 17.623}
            &\tabincell{c}{33.75 \\ 0.915 \\ 9.450}
            &\tabincell{c}{30.68 \\ 0.895 \\ 15.183}
            &\tabincell{c}{33.58 \\ 0.918 \\ 11.969}
            \\
            \midrule
            MST-L 
            & 2.03M
            & 28.15
            &\tabincell{c}{35.40 \\ 0.941 \\ 7.635}
            &\tabincell{c}{35.87 \\ 0.944 \\ 9.396}
            &\tabincell{c}{36.51 \\ 0.953 \\ 6.856}
            &\tabincell{c}{42.27 \\ 0.973 \\ 12.075}
            &\tabincell{c}{32.77 \\ 0.947 \\ 6.842}
            &\tabincell{c}{34.80 \\ 0.955 \\ 10.649}
            &\tabincell{c}{33.66 \\ 0.925 \\ 6.500}
            &\tabincell{c}{32.67 \\ 0.948 \\ 12.889}
            &\tabincell{c}{35.39 \\ 0.949 \\ 8.771}
            &\tabincell{c}{32.50 \\ 0.941 \\ 11.368}
            &\tabincell{c}{35.18 \\ 0.948 \\ 9.308}
            \\
            \midrule
            DAUHST-9stg 
            & 6.15M
            & 79.50
            &\tabincell{c}{37.25 \\ 0.958 \\ 5.946}
            &\tabincell{c}{39.02 \\ 0.967 \\ 7.624}
            &\tabincell{c}{41.05 \\ 0.971 \\ 4.459}
            &\tabincell{c}{46.15 \\ 0.983 \\ 10.837}
            &\tabincell{c}{35.80 \\ 0.969 \\ 5.151}
            &\tabincell{c}{37.08 \\ 0.970 \\ 9.600}
            &\tabincell{c}{37.57 \\ 0.963 \\ 4.471}
            &\tabincell{c}{35.10 \\ 0.966 \\ 10.352}
            &\tabincell{c}{40.02 \\ 0.970 \\ 6.176}
            &\tabincell{c}{34.59 \\ 0.956 \\ 9.812}
            &\tabincell{c}{38.36 \\ 0.967 \\ 7.443}
            \\
            \midrule
            RDLUF-Mix$S^2$ 9stg 
            & 1.89M
            & 115.34
            &\tabincell{c}{37.94 \\ 0.966 \\ 4.975}
            &\tabincell{c}{40.95 \\ 0.977 \\ 6.091}
            &\tabincell{c}{43.25 \\ 0.979 \\ 3.403}
            &\tabincell{c}{47.83 \\ 0.990 \\ 6.362}
            &\tabincell{c}{37.11 \\ 0.976 \\ 4.029}
            &\tabincell{c}{37.47 \\ 0.975 \\ 6.981}
            &\tabincell{c}{38.58 \\ 0.969 \\ 4.073}
            &\tabincell{c}{35.50 \\ 0.970 \\ 7.793}
            &\tabincell{c}{41.83 \\ 0.978 \\ 4.456}
            &\tabincell{c}{35.23 \\ 0.962 \\ 6.645}
            &\tabincell{c}{39.57 \\ 0.974 \\ 5.481}
            \\
            \midrule
            \rowcolor{rouse}
            \bf{DERNN-LNLT 5stg}
            & \bf{0.62M}
            & 45.60
            &\tabincell{c}{37.86 \\ 0.963 \\ 5.019}
            &\tabincell{c}{40.28 \\ 0.976 \\ 6.012}
            &\tabincell{c}{42.69 \\ 0.978 \\ 3.530}
            &\tabincell{c}{47.97 \\ 0.990 \\ 6.360}
            &\tabincell{c}{37.11 \\ 0.975 \\ 4.000}
            &\tabincell{c}{37.23 \\ 0.974 \\ 6.885}
            &\tabincell{c}{37.97 \\ 0.967 \\ 4.152}
            &\tabincell{c}{35.82 \\ 0.971 \\ 7.798}
            &\tabincell{c}{41.93 \\ 0.979 \\ 4.568}
            &\tabincell{c}{34.98 \\ 0.959 \\ 6.891}
            &\tabincell{c}{39.38 \\ 0.973 \\ 5.522}
            \\
            \midrule
            \rowcolor{rouse}
            \bf{DERNN-LNLT 7stg}
            & \bf{0.62M}
            & 63.80
            &\tabincell{c}{37.91 \\ 0.964 \\ 4.915}
            &\tabincell{c}{40.75 \\ 0.978 \\ 5.855}
            &\tabincell{c}{42.95 \\ 0.978 \\ 3.445}
            &\tabincell{c}{47.51 \\ 0.990 \\ 6.398}
            &\tabincell{c}{37.81 \\ 0.978 \\ 3.806} 
            &\tabincell{c}{37.37 \\ 0.975 \\ 6.817}
            &\tabincell{c}{38.49 \\ 0.970 \\ 4.012}
            &\tabincell{c}{35.83 \\ 0.971 \\ 7.738}
            &\tabincell{c}{42.47 \\ 0.980 \\ 4.434}
            &\tabincell{c}{35.04 \\ 0.961 \\ 6.550} 
            &\tabincell{c}{39.61 \\ 0.974 \\ 5.397}
            \\
            \midrule
            \rowcolor{rouse}
            \bf{DERNN-LNLT 9stg}
            & \bf{0.62M}
            & 81.99
            &\tabincell{c}{38.26 \\ 0.965 \\ 4.787}
            &\tabincell{c}{40.97 \\ 0.979 \\ 5.718}
            &\tabincell{c}{43.22 \\ 0.979 \\ 3.400}
            &\tabincell{c}{48.10 \\ 0.991 \\ 6.370}
            &\tabincell{c}{38.08 \\ 0.980 \\ 3.695}
            &\tabincell{c}{37.41 \\ 0.975 \\ 6.753}
            &\tabincell{c}{38.83 \\ 0.971 \\ 3.936}
            &\tabincell{c}{36.41 \\ 0.973 \\ 7.542}
            &\tabincell{c}{42.87 \\ 0.981 \\ 4.335}
            &\tabincell{c}{35.15 \\ 0.962 \\ 6.279}
            &\tabincell{c}{39.93 \\ 0.976 \\ 5.281}
            \\
            \midrule
            \rowcolor{rouse}
            \bf{DERNN-LNLT 9stg$^*$}
            & 1.04M
            & 134.18
            &\tabincell{c}{\bf{38.49} \\ \bf{0.968} \\ \bf{4.570}}
            &\tabincell{c}{\bf{41.27} \\ \bf{0.980} \\ \bf{5.486}}
            &\tabincell{c}{\bf{43.97} \\ \bf{0.980} \\ \bf{3.233}}
            &\tabincell{c}{\bf{48.61} \\ \bf{0.992} \\ \bf{5.879}}
            &\tabincell{c}{\bf{38.29} \\ \bf{0.981} \\ \bf{3.617}}
            &\tabincell{c}{\bf{37.81} \\ \bf{0.977} \\ \bf{6.596}}
            &\tabincell{c}{\bf{39.30} \\ \bf{0.973} \\ \bf{3.846}}
            &\tabincell{c}{\bf{36.51} \\ \bf{0.974} \\ \bf{7.303}}
            &\tabincell{c}{\bf{43.38} \\ \bf{0.983} \\ \bf{4.207}}
            &\tabincell{c}{\bf{35.61} \\ \bf{0.966} \\ \bf{6.008}}
            &\tabincell{c}{\bf{40.33} \\ \bf{0.977} \\ \bf{5.075}}
            \\
		\bottomrule[0.2em]
	\end{tabular}}
	\label{Tab:performance}
\end{table*}

%% file: tables/table2.tex
\begin{table}[!htbp]
    \centering
    \caption{\small Break-down ablation studies of every component. }
    \begin{tabular}{c c c  c c c c}
            \toprule
              & & Params & PSNR & SSIM\\
            \midrule
            1 & Baseline & 2.324M  & 38.04 \color{green}{\bf (+0.00)} & 0.969 \\
                2 & 1 + Non-Local MSA & 2.948M & 38.51 \color{green}{\bf (+0.47)} & 0.970 \\
              3 & 2 + $\Phi^R_k$ & 3.014M & 38.88 \color{green}{\bf (+0.37)} & 0.971 \\
                4 & 3 + $\mu_k$ & 3.112M & 38.95 \color{green}{\bf (+0.07)} & 0.972 \\
                5 & 4 + $\eta_k$ & 3.114M & 39.03 \color{green}{\bf (+0.08)} & 0.972 \\
                6 & 5 + RNN & \bf{0.624M} & \bf 39.38 \color{green}{\bf (+0.35)} & \bf{0.973} \\
            \bottomrule
    \end{tabular}
    \label{tab:Break-down ablation}
\end{table}

%% file: tables/table5.tex
\begin{table}[H]
    \centering
    \caption{\small Comparison with Local MSA and other Local and Non-Local Priors.}
    \scalebox{0.9}{
        \begin{tabular}{c c c  c c c c}
        		\toprule
        		  & & Params & Gflops  & PSNR & SSIM & SAM \ \\
        		\midrule
        		(1) & Local & 2.324M & 31.98 & 38.04 & 0.969 & 4.180 \\
                    (2) & Local + Local & \textbf{2.948M} & \textbf{39.16} & 38.23 & 0.969 & 4.100 \\
        		  (3) & Local + Non-Local & \textbf{2.948M} & \textbf{39.16} & \textbf{38.51} & \textbf{0.970} & \textbf{3.784} \\
                    (4) & HS-MSA & 4.048M & 40.94 & 38.40 & \textbf{0.970} & 3.940 \\
                    (5) & multi-axis gMLPs & 4.091M & 52.01 & 38.12 & 0.969 & 4.035\\
        		\bottomrule
        \end{tabular}
    }
    \label{tab: Ablation}
\end{table}

%% file: tables/table3.tex
\begin{table}[!htbp]
    \centering
    \caption{\small Ablation of numbers of stages.}
    \begin{tabular}{c c c  c c c c}
            \toprule
            Number of stages & Params & PSNR & SSIM\\
            \midrule
            3  & \bf{0.62M} & 38.58 \color{green}{\bf (+0.00)} & 0.969 \\
            5 & \bf{0.62M} & 39.38 \color{green}{\bf (+0.80)} & 0.973 \\
            7 & \bf{0.62M} & 39.61 \color{green}{\bf (+0.23)} & 0.974 \\
            9 & \bf{0.62M} & \bf 39.93 \color{green}{\bf (+0.32)} &\bf 0.976 \\
            $\text{9 wo RM}$ & 5.60M & 39.23 \color{red}{\bf (-0.70)} & 0.972 \\
            \bottomrule
    \end{tabular}
    \label{tab: RM}
\end{table}

%% file: bare_jrnl_new_sample4.bbl
\begin{thebibliography}{10}
\providecommand{\url}[1]{#1}
\csname url@samestyle\endcsname
\providecommand{\newblock}{\relax}
\providecommand{\bibinfo}[2]{#2}
\providecommand{\BIBentrySTDinterwordspacing}{\spaceskip=0pt\relax}
\providecommand{\BIBentryALTinterwordstretchfactor}{4}
\providecommand{\BIBentryALTinterwordspacing}{\spaceskip=\fontdimen2\font plus
\BIBentryALTinterwordstretchfactor\fontdimen3\font minus \fontdimen4\font\relax}
\providecommand{\BIBforeignlanguage}[2]{{%
\expandafter\ifx\csname l@#1\endcsname\relax
\typeout{** WARNING: IEEEtran.bst: No hyphenation pattern has been}%
\typeout{** loaded for the language `#1'. Using the pattern for}%
\typeout{** the default language instead.}%
\else
\language=\csname l@#1\endcsname
\fi
#2}}
\providecommand{\BIBdecl}{\relax}
\BIBdecl

\bibitem{rs1}
M.~Borengasser, W.~S. Hungate, and R.~Watkins, \emph{Hyperspectral remote sensing: principles and applications}.\hskip 1em plus 0.5em minus 0.4em\relax CRC press, 2007.

\bibitem{rs2}
Z.~Zhang, D.~Liu, D.~Gao, and G.~Shi, ``S$^3$net: Spectral--spatial--semantic network for hyperspectral image classification with the multiway attention mechanism,'' \emph{IEEE Transactions on Geoscience and Remote Sensing}, vol.~60, pp. 1--17, 2021.

\bibitem{rs3}
------, ``A novel spectral-spatial multi-scale network for hyperspectral image classification with the res2net block,'' \emph{International Journal of Remote Sensing}, vol.~43, no.~3, pp. 751--777, 2022.

\bibitem{med1}
A.~Bjorgan and L.~L. Randeberg, ``Towards real-time medical diagnostics using hyperspectral imaging technology,'' in \emph{European Conference on Biomedical Optics}.\hskip 1em plus 0.5em minus 0.4em\relax Optica Publishing Group, 2015, p. 953712.

\bibitem{med2}
Z.~Meng, M.~Qiao, J.~Ma, Z.~Yu, K.~Xu, and X.~Yuan, ``Snapshot multispectral endomicroscopy,'' \emph{Optics Letters}, vol.~45, no.~14, pp. 3897--3900, 2020.

\bibitem{industrial}
M.~H. Kim, T.~A. Harvey, D.~S. Kittle, H.~Rushmeier, J.~Dorsey, R.~O. Prum, and D.~J. Brady, ``3d imaging spectroscopy for measuring hyperspectral patterns on solid objects,'' \emph{ACM Transactions on Graphics (TOG)}, vol.~31, no.~4, pp. 1--11, 2012.

\bibitem{arce2013compressive}
G.~R. Arce, D.~J. Brady, L.~Carin, H.~Arguello, and D.~S. Kittle, ``Compressive coded aperture spectral imaging: An introduction,'' \emph{IEEE Signal Processing Magazine}, vol.~31, no.~1, pp. 105--115, 2013.

\bibitem{wagadarikar2008single}
A.~Wagadarikar, R.~John, R.~Willett, and D.~Brady, ``Single disperser design for coded aperture snapshot spectral imaging,'' \emph{Applied optics}, vol.~47, no.~10, pp. B44--B51, 2008.

\bibitem{wang2015high}
L.~Wang, Z.~Xiong, D.~Gao, G.~Shi, W.~Zeng, and F.~Wu, ``High-speed hyperspectral video acquisition with a dual-camera architecture,'' in \emph{Proceedings of the IEEE Conference on Computer Vision and Pattern Recognition}, 2015, pp. 4942--4950.

\bibitem{gap-net}
Z.~Meng, S.~Jalali, and X.~Yuan, ``Gap-net for snapshot compressive imaging,'' \emph{arXiv preprint arXiv:2012.08364}, 2020.

\bibitem{admm-net}
J.~Ma, X.-Y. Liu, Z.~Shou, and X.~Yuan, ``Deep tensor admm-net for snapshot compressive imaging,'' in \emph{Proceedings of the IEEE/CVF International Conference on Computer Vision}, 2019, pp. 10\,223--10\,232.

\bibitem{dgsmp}
T.~Huang, W.~Dong, X.~Yuan, J.~Wu, and G.~Shi, ``Deep gaussian scale mixture prior for spectral compressive imaging,'' in \emph{Proceedings of the IEEE/CVF Conference on Computer Vision and Pattern Recognition}, 2021, pp. 16\,216--16\,225.

\bibitem{DPIR}
K.~Zhang, Y.~Li, W.~Zuo, L.~Zhang, L.~Van~Gool, and R.~Timofte, ``Plug-and-play image restoration with deep denoiser prior,'' \emph{IEEE Transactions on Pattern Analysis and Machine Intelligence}, vol.~44, no.~10, pp. 6360--6376, 2021.

\bibitem{herosnet}
X.~Zhang, Y.~Zhang, R.~Xiong, Q.~Sun, and J.~Zhang, ``Herosnet: Hyperspectral explicable reconstruction and optimal sampling deep network for snapshot compressive imaging,'' in \emph{Proceedings of the IEEE/CVF Conference on Computer Vision and Pattern Recognition}, 2022, pp. 17\,532--17\,541.

\bibitem{dauhst}
Y.~Cai, J.~Lin, H.~Wang, X.~Yuan, H.~Ding, Y.~Zhang, R.~Timofte, and L.~V. Gool, ``Degradation-aware unfolding half-shuffle transformer for spectral compressive imaging,'' \emph{Advances in Neural Information Processing Systems}, vol.~35, pp. 37\,749--37\,761, 2022.

\bibitem{rdluf_mixs2}
Y.~Dong, D.~Gao, T.~Qiu, Y.~Li, M.~Yang, and G.~Shi, ``Residual degradation learning unfolding framework with mixing priors across spectral and spatial for compressive spectral imaging,'' in \emph{Proceedings of the IEEE/CVF Conference on Computer Vision and Pattern Recognition}, 2023, pp. 22\,262--22\,271.

\bibitem{wang2016adaptive}
L.~Wang, Z.~Xiong, G.~Shi, F.~Wu, and W.~Zeng, ``Adaptive nonlocal sparse representation for dual-camera compressive hyperspectral imaging,'' \emph{IEEE transactions on pattern analysis and machine intelligence}, vol.~39, no.~10, pp. 2104--2111, 2016.

\bibitem{zhang2019computational}
S.~Zhang, L.~Wang, Y.~Fu, X.~Zhong, and H.~Huang, ``Computational hyperspectral imaging based on dimension-discriminative low-rank tensor recovery,'' in \emph{Proceedings of the IEEE/CVF International Conference on Computer Vision}, 2019, pp. 10\,183--10\,192.

\bibitem{l-net}
X.~Miao, X.~Yuan, Y.~Pu, and V.~Athitsos, ``l-net: Reconstruct hyperspectral images from a snapshot measurement,'' in \emph{Proceedings of the IEEE/CVF International Conference on Computer Vision}, 2019, pp. 4059--4069.

\bibitem{tsa-net}
Z.~Meng, J.~Ma, and X.~Yuan, ``End-to-end low cost compressive spectral imaging with spatial-spectral self-attention,'' in \emph{European conference on computer vision}.\hskip 1em plus 0.5em minus 0.4em\relax Springer, 2020, pp. 187--204.

\bibitem{swin}
Z.~Liu, Y.~Lin, Y.~Cao, H.~Hu, Y.~Wei, Z.~Zhang, S.~Lin, and B.~Guo, ``Swin transformer: Hierarchical vision transformer using shifted windows,'' in \emph{Proceedings of the IEEE/CVF international conference on computer vision}, 2021, pp. 10\,012--10\,022.

\bibitem{vit}
A.~Dosovitskiy, L.~Beyer, A.~Kolesnikov, D.~Weissenborn, X.~Zhai, T.~Unterthiner, M.~Dehghani, M.~Minderer, G.~Heigold, S.~Gelly \emph{et~al.}, ``An image is worth 16x16 words: Transformers for image recognition at scale,'' \emph{arXiv preprint arXiv:2010.11929}, 2020.

\bibitem{maxim}
Z.~Tu, H.~Talebi, H.~Zhang, F.~Yang, P.~Milanfar, A.~Bovik, and Y.~Li, ``Maxim: Multi-axis mlp for image processing,'' in \emph{Proceedings of the IEEE/CVF Conference on Computer Vision and Pattern Recognition}, 2022, pp. 5769--5780.

\bibitem{nlm}
A.~Buades, B.~Coll, and J.-M. Morel, ``Non-local means denoising,'' \emph{Image Processing On Line}, vol.~1, pp. 208--212, 2011.

\bibitem{mprnet}
S.~W. Zamir, A.~Arora, S.~Khan, M.~Hayat, F.~S. Khan, M.-H. Yang, and L.~Shao, ``Multi-stage progressive image restoration,'' in \emph{Proceedings of the IEEE/CVF conference on computer vision and pattern recognition}, 2021, pp. 14\,821--14\,831.

\bibitem{dgunet}
C.~Mou, Q.~Wang, and J.~Zhang, ``Deep generalized unfolding networks for image restoration,'' in \emph{Proceedings of the IEEE/CVF Conference on Computer Vision and Pattern Recognition}, 2022, pp. 17\,399--17\,410.

\bibitem{pgd}
A.~Beck and M.~Teboulle, ``A fast iterative shrinkage-thresholding algorithm for linear inverse problems,'' \emph{SIAM journal on imaging sciences}, vol.~2, no.~1, pp. 183--202, 2009.

\bibitem{admm}
S.~Boyd, N.~Parikh, E.~Chu, B.~Peleato, J.~Eckstein \emph{et~al.}, ``Distributed optimization and statistical learning via the alternating direction method of multipliers,'' \emph{Foundations and Trends{\textregistered} in Machine learning}, vol.~3, no.~1, pp. 1--122, 2011.

\bibitem{hqs}
R.~He, W.-S. Zheng, T.~Tan, and Z.~Sun, ``Half-quadratic-based iterative minimization for robust sparse representation,'' \emph{IEEE transactions on pattern analysis and machine intelligence}, vol.~36, no.~2, pp. 261--275, 2013.

\bibitem{meinhardt2017learning}
T.~Meinhardt, M.~Moller, C.~Hazirbas, and D.~Cremers, ``Learning proximal operators: Using denoising networks for regularizing inverse imaging problems,'' in \emph{Proceedings of the IEEE International Conference on Computer Vision}, 2017, pp. 1781--1790.

\bibitem{ryu2019plug}
E.~Ryu, J.~Liu, S.~Wang, X.~Chen, Z.~Wang, and W.~Yin, ``Plug-and-play methods provably converge with properly trained denoisers,'' in \emph{International Conference on Machine Learning}.\hskip 1em plus 0.5em minus 0.4em\relax PMLR, 2019, pp. 5546--5557.

\bibitem{yuan2020plug}
X.~Yuan, Y.~Liu, J.~Suo, and Q.~Dai, ``Plug-and-play algorithms for large-scale snapshot compressive imaging,'' in \emph{Proceedings of the IEEE/CVF Conference on Computer Vision and Pattern Recognition}, 2020, pp. 1447--1457.

\bibitem{zhang2017learning}
K.~Zhang, W.~Zuo, S.~Gu, and L.~Zhang, ``Learning deep cnn denoiser prior for image restoration,'' in \emph{Proceedings of the IEEE conference on computer vision and pattern recognition}, 2017, pp. 3929--3938.

\bibitem{zhang2019deep}
K.~Zhang, W.~Zuo, and L.~Zhang, ``Deep plug-and-play super-resolution for arbitrary blur kernels,'' in \emph{Proceedings of the IEEE/CVF Conference on Computer Vision and Pattern Recognition}, 2019, pp. 1671--1681.

\bibitem{dgsmp-pami}
T.~Huang, X.~Yuan, W.~Dong, J.~Wu, and G.~Shi, ``Deep gaussian scale mixture prior for image reconstruction,'' \emph{IEEE Transactions on Pattern Analysis and Machine Intelligence}, 2023.

\bibitem{mst}
Y.~Cai, J.~Lin, X.~Hu, H.~Wang, X.~Yuan, Y.~Zhang, R.~Timofte, and L.~Van~Gool, ``Mask-guided spectral-wise transformer for efficient hyperspectral image reconstruction,'' in \emph{Proceedings of the IEEE/CVF Conference on Computer Vision and Pattern Recognition}, 2022, pp. 17\,502--17\,511.

\bibitem{cst}
------, ``Coarse-to-fine sparse transformer for hyperspectral image reconstruction,'' in \emph{European Conference on Computer Vision}.\hskip 1em plus 0.5em minus 0.4em\relax Springer, 2022, pp. 686--704.

\bibitem{restormer}
S.~W. Zamir, A.~Arora, S.~Khan, M.~Hayat, F.~S. Khan, and M.-H. Yang, ``Restormer: Efficient transformer for high-resolution image restoration,'' in \emph{Proceedings of the IEEE/CVF conference on computer vision and pattern recognition}, 2022, pp. 5728--5739.

\bibitem{mobilenets}
A.~G. Howard, M.~Zhu, B.~Chen, D.~Kalenichenko, W.~Wang, T.~Weyand, M.~Andreetto, and H.~Adam, ``Mobilenets: Efficient convolutional neural networks for mobile vision applications,'' \emph{arXiv preprint arXiv:1704.04861}, 2017.

\bibitem{cave}
F.~Yasuma, T.~Mitsunaga, D.~Iso, and S.~K. Nayar, ``Generalized assorted pixel camera: postcapture control of resolution, dynamic range, and spectrum,'' \emph{IEEE transactions on image processing}, vol.~19, no.~9, pp. 2241--2253, 2010.

\bibitem{kaist}
I.~Choi, M.~Kim, D.~Gutierrez, D.~Jeon, and G.~Nam, ``High-quality hyperspectral reconstruction using a spectral prior,'' Tech. Rep., 2017.

\bibitem{ssim}
Z.~Wang, A.~C. Bovik, H.~R. Sheikh, and E.~P. Simoncelli, ``Image quality assessment: from error visibility to structural similarity,'' \emph{IEEE transactions on image processing}, vol.~13, no.~4, pp. 600--612, 2004.

\bibitem{yolov4}
A.~Bochkovskiy, C.-Y. Wang, and H.-Y.~M. Liao, ``Yolov4: Optimal speed and accuracy of object detection,'' \emph{arXiv preprint arXiv:2004.10934}, 2020.

\bibitem{twist}
J.~M. Bioucas-Dias and M.~A. Figueiredo, ``A new twist: Two-step iterative shrinkage/thresholding algorithms for image restoration,'' \emph{IEEE Transactions on Image processing}, vol.~16, no.~12, pp. 2992--3004, 2007.

\bibitem{gap-tv}
X.~Yuan, ``Generalized alternating projection based total variation minimization for compressive sensing,'' in \emph{2016 IEEE International conference on image processing (ICIP)}.\hskip 1em plus 0.5em minus 0.4em\relax IEEE, 2016, pp. 2539--2543.

\bibitem{desci}
Y.~Liu, X.~Yuan, J.~Suo, D.~J. Brady, and Q.~Dai, ``Rank minimization for snapshot compressive imaging,'' \emph{IEEE transactions on pattern analysis and machine intelligence}, vol.~41, no.~12, pp. 2990--3006, 2018.

\end{thebibliography}
